\documentclass[a4paper,preprintnumbers,showpacs,onecolumn,superscriptaddress,nofootinbib,amsmath,amssymb,notitlepage]{revtex4-1}

\usepackage{enumitem}
\usepackage{times}
\usepackage{dcolumn}
\usepackage{mathrsfs}
\usepackage{hyperref}
\usepackage{amsmath}
\usepackage{mathtools}
\usepackage{bbm}
\usepackage{bm}
\usepackage{amsfonts}
\usepackage{amssymb}
\usepackage{mathrsfs}
\usepackage{wasysym}
\usepackage{graphicx}
\usepackage[]{subfigure}
\usepackage{filecontents}
\usepackage[dvipsnames]{xcolor}

\hypersetup{
    bookmarks=true,         
    pdftoolbar=true,        
    pdfmenubar=true,        
    pdffitwindow=true,     
    pdfstartview={FitH},     
    pdftitle={My title},     
    pdfauthor={author},      
    pdfsubject={Subject},    
    pdfcreator={Creator},    
    pdfproducer={Producer},  
    pdfkeywords={keyword1} 
    pdfnewwindow=true,       
    colorlinks=true ,        
    linkcolor=blue  ,        
    citecolor=blue,      
    filecolor=green,       
    urlcolor=blue            
}
\usepackage{epstopdf}
\usepackage{stackengine}
\stackMath
\newcommand\tsup[2][2]{%
 \def\useanchorwidth{T}%
  \ifnum#1>1%
    \stackon[-1.3ex]{\tsup[\numexpr#1-1\relax]{#2}}{\mathchar"307E}%
  \else%
    \stackon[-1ex]{#2}{\mathchar"307E}%
  \fi%
}

\newcommand{\arccosh}{\operatorname{arccosh}}
\newcommand{\sech}{\operatorname{sech}}

\newcommand{\ed}{\mathrm{d}}
\newcommand{\mA}{\mathcal{A}}
\newcommand{\mB}{\mathcal{B}}

\newcommand{\mP}{\mathcal{P}}

\begin{document}

\title{Analytical study of light ray trajectories in Kerr spacetime in the presence of an inhomogeneous anisotropic plasma}

\author{Mohsen Fathi}
\email{mohsen.fathi@postgrado.uv.cl}
\affiliation{Instituto de F\'{i}sica y Astronom\'{i}a, Universidad de Valpara\'{i}so,
Avenida Gran Breta\~{n}a 1111, Valpara\'{i}so, Chile}

\author{Marco Olivares}
\email{marco.olivaresr@mail.udp.cl}
\affiliation{Facultad de Ingenier\'{i}a y Ciencias, Universidad Diego Portales, Avenida Ej\'{e}rcito Libertador 441, Santiago, Chile}

\author{J.R. Villanueva}
\email{jose.villanueva@uv.cl}
\affiliation{Instituto de F\'{i}sica y Astronom\'{i}a, Universidad de Valpara\'{i}so,
Avenida Gran Breta\~{n}a 1111, Valpara\'{i}so, Chile}

\date{\today}

\begin{abstract}
We calculate the exact solutions to the equations of motion that govern the light ray trajectories as they travel in a Kerr black hole's exterior that is considered to be filled with an inhomogeneous and anisotropic plasmic medium. 
This is approached by characterizing the plasma through conceiving a radial and an angular structure function, which are let to be constant. The description of the motion is carried out by using the Hamilton-Jacobi method, that allows defining two effective potentials, characterizing the evolution of the polar coordinates. The elliptic integrals of motion are then solved analytically, and the evolution of coordinates is expressed in terms of the Mino time. This way, the three-dimensional demonstrations of the light ray trajectories are given respectively.
\bigskip

{\noindent{\textit{keywords}}: Light ray trajectories, Kerr black hole, anisotropic plasma}\\

\end{abstract}

\maketitle

\section{Introduction}\label{sec:introduction}

The analysis of light ray trajectories in the spacetimes associated with  black holes, dates back to the very early years of general relativity. However, it was not until 1930, that a reliable analytical derivation of particle trajectories in Schwarzschild spacetime was presented by Hagihara \cite{Hagihara:1930}. Since then, efforts for the analytical determination of the gravitational lensing caused by black holes have been continued constantly, and significant formulations have been derived, which also include ways of confining the black hole shadow \cite{Bardeen:1972a,Bardeen:1973a,Bardeen:1973b,Chandrasekhar:2002}. The importance of analytical treatment of particle geodesics in black hole spacetimes, is to provide the possibility to do a strict comparison of the theoretical predictions with the observational data. This kind of treatment is, therefore, of interest among respected scientists. For example, the study of gravitational lensing was resurrected by the very famous paper  \cite{Virbhadra:2000} in Schwarzschild spacetime, which founded many similar researches specified to the case of rotating black holes. The non-static black hole spacetimes, however, correspond to complicated equations of motion that necessitate more advanced mathematical methods to be established. In fact, the search for such methods have led to numerous publications. Along the same efforts, and for the case of the application of elliptic integrals in the calculation of light ray geodesics (which is also the interest of this work), several criteria have been discussed and resolved in Refs.~\cite{{kraniotis_precise_2004,beckwith_extreme_2005,kraniotis_frame_2005,hackmann_complete_2008,hackmann_geodesic_2008,bisnovatyi-kogan_strong_2008,kagramanova_analytic_2010,hackmann_analytical_2010,hackmann_complete_2010,enolski_inversion_2011,kraniotis_precise_2011,enolski_inversion_2012,gibbons_application_2012,munoz_orbits_2014,kraniotis_gravitational_2014,de_falco_approximate_2016,soroushfar_detailed_2016,barlow_asymptotically_2017,uniyal_null_2018,villanueva_null_2018,chatterjee_analytic_2019,hsiao_equatorial_2020,gralla_null_2020,hendi_simulation_2020,fathi_classical_2020,kraniotis_gravitational_2021}}. Furthermore, following the discussions given by Synge in Ref.~\cite{Synge:1960}, such mathematical methods and similar ones, have made possible the determination and confinement of light propagation in the spacetimes of theoretical (non-)static black holes that are surrounded by plasmic or dark fluid media \cite{bisnovatyi-kogan_gravitational_2009,bisnovatyi-kogan_gravitational_2010,tsupko_gravitational_2013,morozova_gravitational_2013,bisnovatyi-kogan_gravitational_2015,perlick_influence_2015,atamurotov_optical_2015,abdujabbarov_shadow_2016,bisnovatyi-kogan_gravitational_2017,perlick_light_2017,schulze-koops_sachs_2017,abdujabbarov_shadow_2017,liu_effects_2017,haroon_shadow_2019,kimpson_spatial_2019,babar_optical_2020,junior_spinning_2020,Badia:2021}. Despite this, the analytical treatment of light ray trajectories in non-vacuum black hole surroundings is seemed to be overlooked. In fact, such analysis can help classifying the other possible orbits in such conditions, that are not usually considered to determine the black holes' photon spheres and shadows. Hence, in this paper, we apply the aforementioned mathematical methods to calculate the light ray paths while they travel in a non-vacuum non-static spacetime. 

Accordingly, we investigate the optical nature of the exterior spacetime of a Kerr black hole, considering that it is immersed in an inhomogeneous anisotropic plasmic medium whose structural functions are constant. To achieve this objective, in Sec. \ref{sec:HJEOM}, we use the Hamilton-Jacobi formalism to obtain the basic relations that allow us to describe the propagation of light in a dispersive medium. In order to get a better perception of the equations of motion, the Mino time is introduced as the trajectory parameter. In particular, in Sec. \ref{sec:eqofmot}, we use simple forms for the structural functions, i.e. $f_r$ and $f_{\theta}$, that make it possible obtaining exact analytical solutions to the equations of motion, which in this paper, are expressed in terms of the Weierstra${\ss}$ian elliptic functions. With the aim of the Mino time as the trajectories' curve parameter, we obtain the spatial three-dimensional representations of these exact solutions. We conclude in Sec. \ref{sec:concl}. Throughout this work we will use the geometric units, in which $G=c=\hbar=1$.

\section{Light propagation in an inhomogeneous plasma in Kerr spacetime}
\label{sec:HJEOM}

To study the light propagation in a medium, we need to consider the phase space dynamics, by introducing an appropriate Hamiltonian. For light rays propagating in a non-magnetized pressure-less plasma, this Hamiltonian is characterized as
\begin{equation}\label{H1}
\mathcal{H}(\bm{x},\bm{p})=\frac{1}{2}\left[g^{\mu \nu}(\bm{x})p_{\mu}p_{\nu}+
\omega_p^2(\bm{x})\right],
\end{equation}
 in the coordinate system $\bm{x}\equiv x^{\mu}=(x^0,x^1,x^2,x^3)$, where $\bm{p}\equiv p_{\mu}=(p_0,p_1,p_2,p_3)$ is the canonical momentum covector. Here, $\omega_p(\bm{x})$ is the plasma electron frequency given by (for a review see Ref.~\cite{Kogan:2017})
\begin{equation}\label{eq:omegap}
    \omega_p(\bm{x}) = \frac{4 \pi e^2}{m_e} N_p(\bm{x}),
\end{equation}
in terms of the electron charge $e$, electron mass $m_e$, and the number density $N_p$. Defining the plasmic refractive index \cite{Atamurotov:2015}
\begin{equation}\label{eq:rn2}
    n^2 = 1 + \frac{p_\mu p^\mu}{(p_\nu u^\nu)^2}
    = 1-\frac{\omega_p^2}{\omega^2},
\end{equation}
with $\bm{u}$ as the observer's four-velocity, we can recast the Hamiltonian as
\begin{equation}\label{eq:hamilton_1}
    \mathcal{H}(\bm{x},\bm{p}) = \frac{1}{2}\left[
    g^{\mu\nu}(\bm{x}) p_\mu p_\nu - (n^2-1)(p_\nu u^\nu)^2
    \right].
\end{equation}
Here, the quantity $p_\nu u^\nu = -\omega$ gives the effective energy of the photons of frequency $\omega$, as measured by the observer. This way, the refractive index $n\equiv n(\Vec{x},\omega)$ constructs the optical metric $\mathfrak{o}^{\mu\nu}$ of the plasmic medium, such that $\mathcal{H}=\frac{1}{2}\mathfrak{o}^{\mu\nu} p_\mu p_\nu$. 
In fact, the Hamilton-Jacobi approach, requires the contribution of the Jacobi action $\mathcal{S}$, given by
\begin{subequations}\label{eq:StoP}
\begin{align}
  &  p_\mu = \frac{\partial\mathcal{S}}{\partial x^\mu},\label{eq:StoP_1}\\
  &  \mathcal{H} = -\frac{\partial\mathcal{S}}{\partial\lambda},\label{eq:StoP_2}
\end{align}
\end{subequations}
with $\lambda$ as the curve parametrization. Now, if the observer is located on the $x^0$ curves, it has the four-velocity $u^\mu = {\delta_0^\mu}/{\sqrt{-g_{00}}},$ which, employing Eqs.~\eqref{eq:StoP}, yields the Hamilton-Jacobi equation as \cite{Atamurotov:2015,Perlick:2017}
\begin{eqnarray}\label{eq:hamilton_2}
    \frac{\partial\mathcal{S}}{\partial\lambda} &=&  -\frac{1}{2}\left[
    g^{\mu\nu} \frac{\partial\mathcal{S}}{\partial x^\mu} \frac{\partial\mathcal{S}}{\partial x^\nu} +\omega_p^2
    \right]\nonumber\\ 
    &=&-\frac{1}{2}\left[
    g^{\mu\nu} \frac{\partial\mathcal{S}}{\partial x^\mu} \frac{\partial\mathcal{S}}{\partial x^\nu} - \left(n^2-1\right)\omega^2
    \right],
\end{eqnarray}
in which, we have used the identity $p_\nu u^\nu = p_0/\sqrt{g_{00}} = -\omega$. Therefore, the general equations governing the light ray trajectories are
\begin{subequations}\label{EqsH}
\begin{align}
&\frac{\partial \mathcal{H}}{\partial p_{\mu}}={\ed x^{\mu}\over \ed\lambda},\label{EqsH1}\\
&\frac{\partial \mathcal{H}}{\partial x^{\mu}}=-{\ed p_{\mu}\over \ed\lambda},\label{EqsH2}\\
&\mathcal{H}=0\label{EqsH3}.
\end{align}
\end{subequations}

\subsection{Light propagating in Kerr spacetime}

The Kerr black hole spacetime is described by the line element
\begin{equation}\label{metr}
\mathrm{d}s^2 = -\left(1-\frac{2 M r}{\rho^2}\right)\ed t^2 + \frac{\rho^2}{\Delta}\ed r^2+\rho^2 \ed\theta^2
+\sin^2\theta \left(r^2+a^2+\frac{2M r a^2 \sin^2\theta}{\rho^2}\right)\ed\phi^2\\
-\frac{4M r a \sin^2\theta}{\rho^2}\, \ed t \ed\phi,
\end{equation}
which, here, is given in terms of the Boyer–Lindquist coordinates $x^\mu=(t, r, \theta, \phi)$. In the above line element
\begin{subequations}\label{eq:rho,Delta}
\begin{align}
    & \Delta = r^2+a^2-2 M r,\label{eq:Delta}\\
    & \rho^2 = {r^2 + a^2 \cos^2 \theta},\label{eq:rho}
\end{align}
\end{subequations}
with $M$ and $a = J/M$ being, respectively, the mass and the spin parameter of the black hole, where $J$ is the black hole's angular momentum. The Kerr black hole spacetime admits for a Cauchy and an event horizon (notated respectively by $r_-$ and $r_+$), whose surfaces are determined by solving $\Delta = 0$, and are given by 
\begin{equation}\label{eq:rH}
    r_\mp=M\mp\sqrt{M^2-a^2}.
\end{equation}
The other significant hypersurface in Kerr spacetime, where $g_{tt} = 0$, corresponds to a region, inside which, no static observes can exist and stationary observers are in the state of $corotation $ with the black hole. This surface of static limit is located at
\begin{equation}\label{eq:rSL}
    r_{\mathrm{SL}}(\theta) = M+\sqrt{M^2-a^2\cos^2\theta},
\end{equation}
and together with the event horizon, forms the black hole's ergosphere in the region $r_+<r<r_{\mathrm{SL}}$.
Throughout this paper, we restrict our study to the domain of outer communications, i.e., the domain outside the event horizon ($r > r_+$), and we consider the case of $a^2\leq M^2$, that corresponds to a black hole rather than a naked singularity. 
The famous method of separation of the Jacobi action, is based on the definition \cite{Carter:1968,Chandrasekhar:2002}
\begin{equation}\label{eq:S_sep}
    \mathcal{S} = -E t + L \phi + \mathcal{S}_r(r) + \mathcal{S}_\theta(\theta) + \frac{1}{2} m^2 \tau, 
\end{equation}
in which $E$, $L$ and $m$ are, respectively, the energy, angular momentum, and the mass associated with the test particles (here, $m=0$). The first two parameters are known as the constants of motion which are specified by the Hamilton equations. Along with the method of separation of the Hamilton-Jacobi equation introduced in Ref.~\cite{perlick_light_2017}, we recast the Hamiltonian (\ref{H1}) in the Kerr spacetime as
\begin{equation}\label{H2}
\mathcal{H}(\bm{x},\bm{p}) = {1\over 2\rho^2}\left[  \Delta \,p_r^2+\,p_{\theta}^2+\left(a p_t\sin \theta+{p_{\phi} \over\sin \theta} \right) ^2\right. 
- \left.{1\over\Delta}\left(p_t(r^2+a^2)
+ap_{\phi}\right)^2+\rho^2 \omega_p^2\right].
\end{equation}
Assuming $\omega_p\equiv\omega_p(r, \theta)$, it is then straightforward to see that $\partial \mathcal{H}/ \partial t=0=\partial \mathcal{H}/ \partial \phi=0$. Therefore, taking into account Eqs.~\eqref{eq:StoP_1} and \eqref{eq:S_sep}, we can specify the constants of motion as
\begin{subequations}\label{ctesmov}
\begin{align}
& E = -\frac{\partial\mathcal{S}}{\partial t} = - p_t = \omega_0,\\
& L =\frac{\partial\mathcal{S}}{\partial\phi} = p_{\phi}.
\end{align}
\end{subequations}
Physically, the angular momentum component $p_{\phi}$ corresponds, to the axial symmetry of the spacetime. The nature of $\omega_0$, on the other hand, becomes clear if we specify the light rays' frequency. In fact, the special case of  $\omega_p(r,\theta)$ corresponds to the three constants of motion, $\mathcal{H}=0$, $\omega_0$  and $L$. Accordingly, we can apply the Carter's method of separation of the Hamilton-Jacobi equation, through which, the light ray trajectories are given in terms of integrable equations \cite{Carter:1968,Chandrasekhar:2002}.  Using Eqs.~(\ref{H2}) and (\ref{ctesmov}), the Hamilton-Jacobi equation, $ \mathcal{H}(\bm{x},\bm{p})=0$, yields
\begin{equation}\label{H3}
0 =  \Delta \,p_r^2+\,p_{\theta}^2+\left(\omega_0\,a\sin \theta-{L \over\sin \theta} \right) ^2 
 - {1\over\Delta}\left[\omega_0(r^2+a^2)
-aL\right]^2+\rho^2\omega_p^2.
\end{equation}
The separability property of the Hamilton-Jacobi equation, demands that Eq.~(\ref{H3}) can be divided into separated $r$-dependent and $\theta$-dependent segments. In Ref.~\cite{perlick_light_2017}, this has been made possible by defining
\begin{equation}
\omega_p(r,\theta)^2= {f_r(r)+f_{\theta}(\theta)\over r^2+a^2\cos^2\theta},
\label{plasma2}
\end{equation}
for some functions $f_{r}(r)$ and $f_{\theta}(\theta)$. Now, the identity 
\begin{equation}
\left(\omega_0\,a\sin \theta-{L \over\sin \theta} \right) ^2=
(L^2\csc^2\theta-a^2\omega_0^2)\cos^2\theta\\
+(L-a\omega_0)^2,
\end{equation}
together with Eqs.~\eqref{H3} and \eqref{plasma2}, separates the Hamilton-Jacobi equation as
\begin{eqnarray}\label{H4}
\mathscr{Q}&=&  \,p_{\theta}^2+(L^2\csc^2\theta-a^2\omega_0^2)\cos^2\theta+f_{\theta}(\theta)\nonumber \\
&=&- \Delta \,p_r^2+{1\over\Delta}\left[\omega_0(r^2+a^2)
-aL\right]^2-(L-a\omega_0)^2-f_{r}(r),
\end{eqnarray}
in which, $\mathscr{Q}$ is the so-called {\it Carter's constant}. Recasting the above equations, yields
\begin{eqnarray}
p_{\theta}^2 &=& \mathscr{Q}-(L^2\csc^2\theta-a^2\omega_0^2)\cos^2\theta-f_{\theta}(\theta),\label{H5}\\
\Delta \,p_r^2 &=& {1\over\Delta}\left[\omega_0(r^2+a^2)
-aL\right]^2-\mathscr{Q}-(L-a\omega_0)^2
-f_{r}(r).\label{H5an}
\end{eqnarray}
Insertion of Eqs. (\ref{H5}) and (\ref{H5an}) into the Hamiltonian  (\ref{H2}), and then using Eq.~\eqref{EqsH1}, provides the first order differential equations of motion as
\begin{eqnarray}\label{basiceqs}
&& \rho^2\frac{\ed r}{\ed\lambda}=\sqrt{\mathcal{R}(r)},\label{basiceqsr}\\ &&\rho^2\frac{\ed\theta}{\ed\lambda}=\sqrt{\Theta (\theta)},\label{basiceqstheta}\\
&&\rho^2\frac{\ed\phi}{\ed\lambda}={ L(\rho^2-2Mr)\csc^2\theta+2Ma\omega_0r\over \Delta}, \label{basiceqsphi} \\
&&\rho^2\frac{\ed t}{\ed\lambda}={ \omega_0\Sigma^2-2MaLr\over \Delta}, \label{basiceqst}
\end{eqnarray}
where
\begin{subequations}\label{basiceqs2}
\begin{align}
& \mathcal{R}(r) = \left[\omega_0( r^2+a^2) -aL\right] ^2-\Delta\left[ \mathscr{Q}+f_r(r)+(L-a\omega_0)^2\right],\label{basiceqs2b}\\
& \Theta(\theta) = \mathscr{Q}-f_{\theta}(\theta)-\cos^2\theta\left(L^2\csc^2\theta-a^2\omega_0^2\right),\label{basiceqs2c} \\
& \Sigma^2=\rho^2\left( r^2+a^2\right) +2Mra^2\sin^2\theta.\label{basiceqs2d}
\end{align}
\end{subequations}
Finally, defining the {dimension-less} \textit{Mino time}\footnote{{Note that, in the geometric units we use, the parameter time has the dimensions of length. Same holds for the curve parameter $\lambda$.}}, $\gamma$, as $\rho^2\ed\gamma=M\ed\lambda$ \cite{Mino:2003},
the equations of motion are now rewritten as
\begin{eqnarray}
\label{basiceqsR}
M{\ed r\over \ed \gamma}&=&\sqrt{\mathcal{R}(r)},\\ 
\label{basiceqstheta}
M{\ed\theta\over \ed \gamma}&=&\sqrt{\Theta (\theta)},\\
\label{basiceqsphi}
M{\ed\phi\over \ed\gamma}&=&{L(\rho^2-2Mr)\csc^2\theta+2Ma\omega_0r\over \Delta}, \\
\label{basiceqste}
M{\ed t\over \ed \gamma}&=&{ \omega_0\Sigma^2-2MaLr\over \Delta}. 
\end{eqnarray}
In the next section, we continue our discussion by analyzing the above equations of motion, in order to find analytically exact solutions to the light ray trajectories travelling in an 
inhomogeneous anisotropic plasma around the black hole.

\section{General analysis of the equations of motion}
\label{sec:eqofmot}

The separation condition in Eq.~\eqref{plasma2} characterizes the plasma, regarding its geometric distribution in the spacetime surrounding the black hole. Therefore, the characteristic functions $f_r(r)$ and $f_\theta(\theta)$, play important roles in defining the plasma's configuration. In this section, we confine the light rays to an inhomogeneous anisotropic plasma, by adopting the particular choices
\begin{eqnarray}
&& f_r(r)\equiv f_r=\Omega_0^2\,R^2 = \mathrm{constant},\label{H61}\\
&& f_{\theta}(\theta)\equiv f_\theta=\Omega_0^2\,a^2=\mathrm{constant.}\label{H62}
\end{eqnarray} 
This way, Eq.~\eqref{plasma2} can be recast as
\begin{equation}
\omega_p(r,\theta)^2 = \Omega_0^2\left({R^2+a^2\over r^2+a^2\cos^2\theta}\right),
\label{plasma4}
\end{equation}
where $R$ is the mean radius of the gravitating object, and $\Omega_0$ is a positive constant. Within the text, we use, frequently, the conventions
\begin{eqnarray}
&& \xi={L\over \omega _0},\label{eq:xi}\\
&& \eta= {\mathscr{Q}\over\omega _0^2},\label{eq:eta}
\end{eqnarray}
in order to simplify the analysis. In what follows, the temporal evolution of the spacetime coordinates are analyzed separately for this plasma, and the relevant exact solutions are presented. 

\subsection{The evolution of the radial distance (the $r$-motion)}

\begin{figure}[t]
\begin{center}
\includegraphics[width=6 cm]{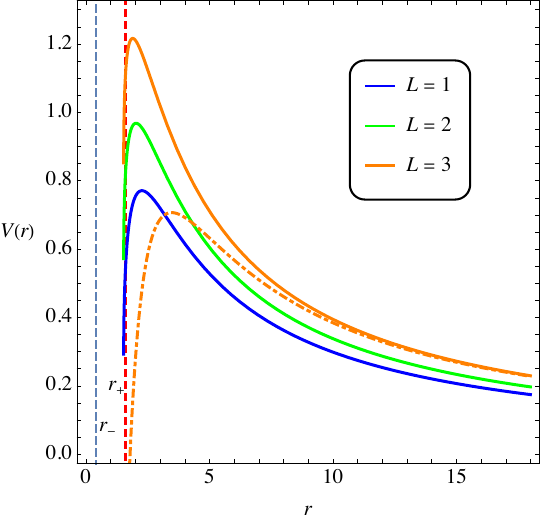}~(a)
\includegraphics[width=6 cm]{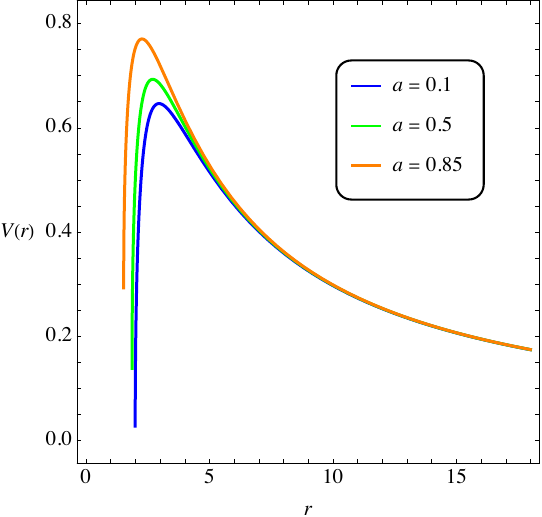}~(b)
\end{center}
\caption{The radial effective potential in an inhomogeneous anisotropic plasma in the region $r>r_+$, plotted for $\mathscr{Q}=9M^2$ and $f_r = 1 M^2$. The diagrams have been generated for (a) $a = 0.85 M$ and three different values of $L$ (the solid lines), and (b) $L = 1 {M}$ and three different values of $a$. The dot-dashed curve in the diagram (a), corresponds to the retrograde motion, which has been plotted for $a = -0.85 M$ and $L = 3 {M}$
({{For these diagrams and for all the forthcoming ones within this paper, the unit along the axes is considered to be ${M}$}}). }
\label{fig:EffectivePotential_general}
\end{figure}

In the study of particle trajectories in curved spacetimes, it is of crucial importance to know how the particles approach and recede from the source of gravity. Based on the nature of the interactions, this study is traditionally done by calculating the radial, effective gravitational potential, that acts on the particles \cite{Misner:1973}. Hence, to scrutinize the $r$-motion for the light ray trajectories in the context under consideration, we focus on the radial equation of motion \eqref{basiceqsR}, and rewrite the expression in Eq.~\eqref{basiceqs2b} as  
\begin{equation}\label{i.12}
    \mathcal{R}(r) = \mathscr{P}(r)\big[\omega_0-V_-(r)\big]\big[\omega_0-V_+(r)\big],
\end{equation}
where $\mathscr{P}(r)=r^4+a^2r^2+2Ma^2r$, and the radial gravitational potentials are given by
\begin{equation}\label{i.13}
V_{\mp}(r) = \frac{1}{\mathscr{P}(r)}\Big\{2 M a L r\\
\mp\Big[\Delta\mathscr{P}(r)\left(\mathscr{Q}+f_r+L^2-\frac{a^2L^2}{\Delta}\right)\\
 +4a^2L^2M^2r^2\frac{}{}\Big]^{\frac{1}{2}}\Big\},
\end{equation}
taking into account the condition \eqref{H61}. The negative branch is not of our interest, since it has no classical interpretations\footnote{In the context of quantum theory of fields, negative energies are related to antiparticles that move backwards in time (see for example Ref.~\cite{Lancaster:2014}).}. We therefore, choose the positive branch of Eq.~\eqref{i.13} as the effective potential, i.e.
$V_{\mathrm{eff}}=V_{+}(r)\equiv V(r)$, noting that $V(r\rightarrow \infty)=0$. In Fig.~\ref{fig:EffectivePotential_general}, this effective potential has been demonstrated for different values of $a$ and $L$. 
According to the diagrams, no stable orbits are expected since the effective potentials do not possess any minimums. Note that, the photons can also pursue a motion along the opposite direction of the black hole's spin, which is the significance of the retrograde motion. As seen in the left panel of Fig.~\ref{fig:EffectivePotential_general}, the effective potential corresponding to the retrograde motion (plotted for $a<0$), exhibits a lower maximum energy for the the same angular momentum. Therefore, photons on the retrograde motion encounter a remarkably smoother gravitational potential.

The possible trajectories are then categorized based on the turning points. However, before proceeding with the determination of the turning points, let us rewrite Eq.~\eqref{basiceqsR} as
\begin{equation}\label{tl11a}
M{\ed r\over \ed \gamma}=\omega_0\sqrt{\mathcal{P}(r)},
\end{equation}
in terms of the characteristic polynomial
\begin{equation}\label{eq:P(r)_char}
    \mathcal{P}(r) =  r^{4}+\mathcal{A} r^2+\mathcal{B} r+\mathcal{C},
\end{equation}
where
\begin{subequations}\label{eq:C12}
\begin{align}
  &  \mathcal{A} = a^2-\xi^2-\eta-\eta_r,\label{c12_A}\\
  &  \mathcal{B} = 2M\left[\eta+\eta_r+\left( \xi-a\right)^2\right],\label{c12_B}\\
  &  \mathcal{C} =-a^2(\eta+\eta_r),\label{c12-C}
\end{align}
\end{subequations}
and $\eta_r= f_r/ \omega _0^2$.

\subsubsection{The turning points}

\begin{figure}[t]
\begin{center}
\includegraphics[width=6cm]{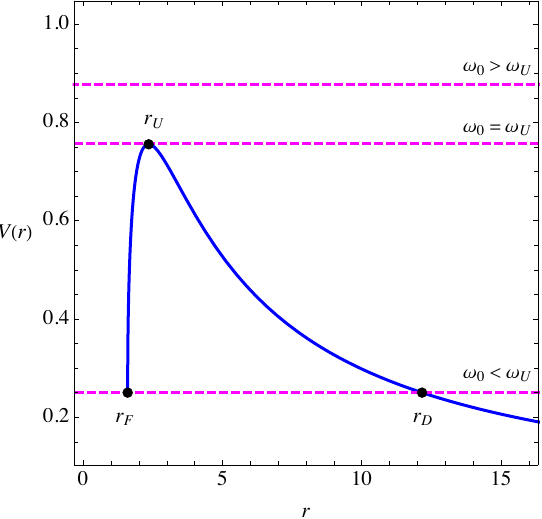}
\end{center}
\caption{A typical effective potential, plotted for $a=0.8 M$, $L = 1 {M}$, and $\mathscr{Q}+f_r = 10 {M^2}$ ({which are the values that will be taken into account for all the forthcoming diagrams in this paper}). The categorization of orbits is done by comparing the photon energy $\omega_0$ with that for photons on the UCO, i.e. $\omega_U$ (for the above values, $\omega_U \approx 0.755232$). {Photons with the energy $\omega_0>\omega_U$, will experience an inevitable fall onto the black hole's event horizon}. For the special case of $\omega_0<\omega_U$, the photons encounter two turning points $r_D$ and $r_F$.}
\label{fig:EffectivePotential_typical}
\end{figure}

\begin{figure}[t]
\begin{center}
\includegraphics[width=5.5cm]{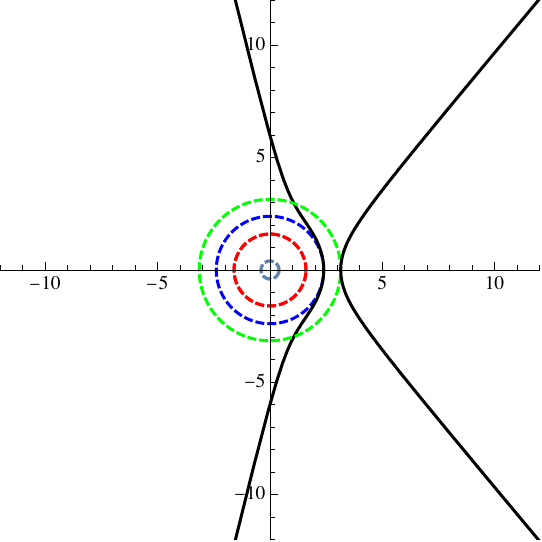}~(a)
\includegraphics[width=5.5cm]{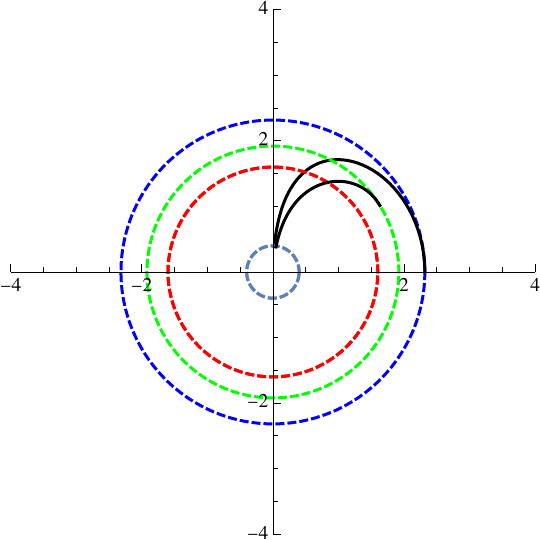}~(b)
\end{center}
\caption{{The polar plots of $r(\gamma)$, respecting (a) 
DFKs and (b) 
DSKs, plotted for ${\omega_0 = 0.755}$ (blue circles), and $\omega_0 = 0.70$ (green circles).
The two inner circles indicate $r_+$ and $r_-$.}}
\label{fig:rdrf}
\end{figure}

\begin{figure}[t]
\begin{center}
\includegraphics[width=7.5cm]{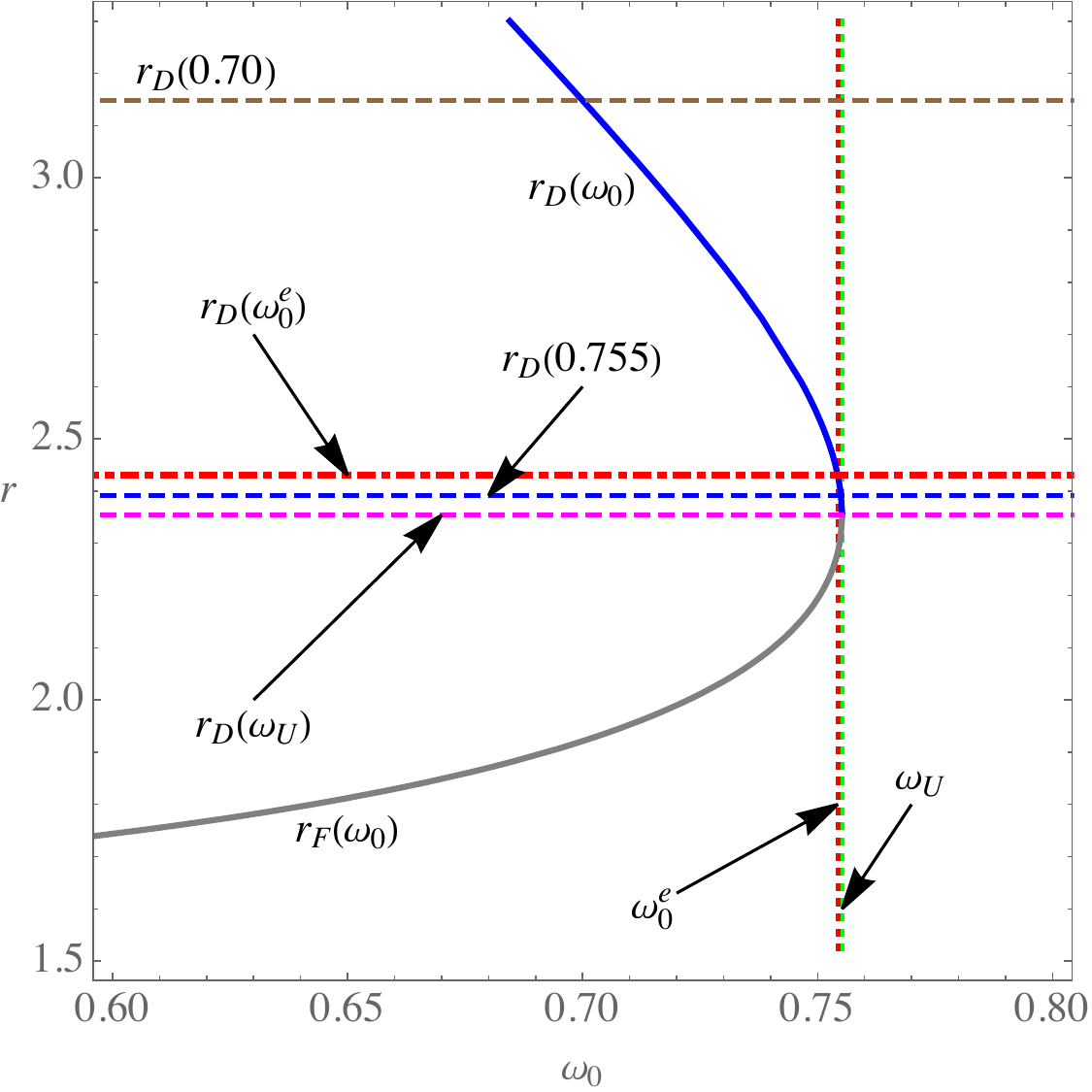}
\end{center}
\caption{{The plot of $r_D(\omega_0)$ and $r_F(\omega_0)$, in accordance to the values given in Fig.~\ref{fig:rdrf}. The dashed lines indicate $r_D(0.70)$, $r_D(0.755)$ and $r_D(\omega_U)$, and the switching radius $r_D(\omega_0^\mathrm{e})$, shown by the dot-dashed line, corresponds to the starting point of a region, where the trajectories begin to change their deflecting character. In this case, $\omega_0^\mathrm{e}\approx 0.7543$.}}
\label{fig:rDomega0}
\end{figure}

\begin{figure}[t]
\begin{center}
\includegraphics[width=5.0cm]{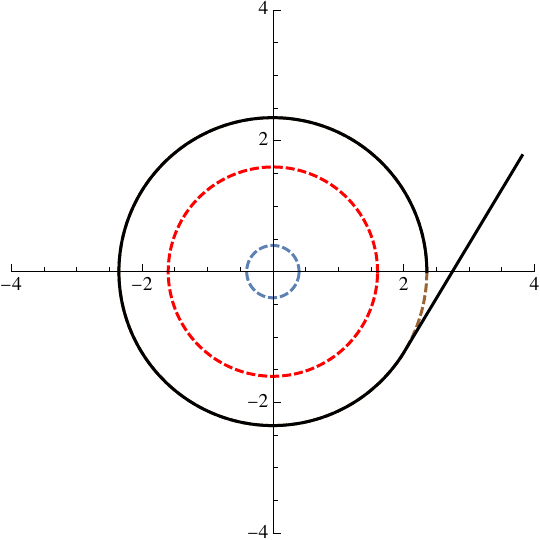}~(a)
\includegraphics[width=5.0cm]{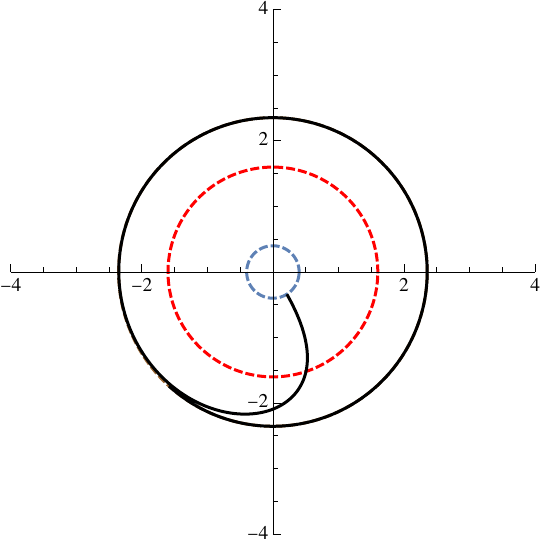}~(b)
\end{center}
\caption{{The polar plots of (a) the UCOFK and (b) the UCOSK, plotted for $\omega_0 \approx \omega_U$. The outer circle, indicates $r_U\approx 2.166051$.}}
\label{fig:ru}
\end{figure}

Let us consider a typical effective potential as demonstrated in Fig.~\ref{fig:EffectivePotential_typical}. The possible types of motion are categorized regarding the photon frequency (energy) $\omega_0$, compared with its value $\omega_U$, for photons on the unstable circular orbits (UCO).
When $\omega_0>\omega_U$, the characteristic polynomial $\mathcal{P}(r)$ has four complex roots, which for $\omega_0=\omega_U$, reduces to two complex roots and a (degenerate) positive real root. This latter corresponds to the UCO at the orbital radius $r_U$. For the case of $\omega_0<\omega_U$, the polynomial has two complex and two positive reals roots, $r_D$ and $r_F$, that correspond to the turning points of the photon orbits. {In fact, Eqs.~(\ref{basiceqsR}) and \eqref{basiceqste} also inform about the characteristics of the turning points. Defining the coordinate velocity $v_c(r) = \ed r/\ed t$, then the turning points, $r_t$, are where $v_c(r_t) = 0$.} Based on the definition given in Eq.~\eqref{tl11a}, this condition is equivalent to $\mathcal{P}(r_t) = 0$, which together with Eq.~\eqref{eq:P(r)_char}, results in the two radii 
\begin{eqnarray}\label{eq:rFrD}
r_D &=& \tilde{R}+\sqrt{\tilde{R}^2-\tilde{Z}},\label{eq:rD}\\
r_F &=& \tilde{R}-\sqrt{\tilde{R}^2-\tilde{Z}},\label{eq:rF}
\end{eqnarray}
where
\begin{subequations}\label{eq:R,Z}
\begin{align}
    &  \tilde{R} = \sqrt{\Xi-{\mathcal{A}\over 6}},\\
    &  \tilde{Z} = 2\tilde{R}^2 +{\mathcal{A}\over 2}+{\mathcal{B}\over 4\tilde{R}},
\end{align}
\end{subequations}
and
\begin{equation}\label{eq:Xi}
    \Xi = 2\sqrt{{\chi_2\over 3}} \cosh\left(\frac{1}{3}\arccosh\left({3\over 2}\chi_3\sqrt{{3\over \chi_2^3}}  \right) \right),
\end{equation}
in which,
\begin{subequations}\label{eq:Chi2,3}
\begin{align}
& \chi_2 = {\mathcal{A}^2\over 48}+{\mathcal{C} \over 4}, \\
& \chi_3 = {\mathcal{A}^3\over 864}+{\mathcal{B}^2\over 64}-{\mathcal{A}\,\mathcal{C} \over 24}.
\end{align}
\end{subequations}
The turning points $r_D$ and $r_F$ correspond to different fates for the trajectories. Respecting Fig.~\ref{fig:EffectivePotential_typical}, photons that approach the black hole at $r_D$, will deflect to infinity by pursuing a hyperbolic type of motion ({deflection of the first kind (DFK)}). The equation of motion \eqref{tl11a}, if solved at the vicinity of the deflection point $r_D$, yields the DFK as (appendix \ref{app:D})
\begin{equation}\label{eq:r(gamma)_def}
    r_d(\gamma) = \frac{\left[1+u_d(\gamma) \right] r_D}{u_d(\gamma)},
\end{equation}
in which
\begin{equation}\label{eq:u(gamma)_def}
    u_d(\gamma) = 4\wp\left(\frac{\omega_0\sqrt{C_3}\,\gamma}{M r_D}\right)-\frac{\alpha_1}{3},
\end{equation}
where $\wp(x)\equiv\wp(x;\tilde{g}_2,\tilde{g}_3)$ is the elliptic Weierstra$\ss$ian $\wp$ function (see appendix \ref{app:A}), here, associated with the invariants
\begin{subequations}\label{eq:tg2tg3}
\begin{align}
    & \tilde{g}_2 = \frac{1}{4}\left(
    \frac{\alpha_1^2}{3}-\alpha_2
    \right),\\
    & \tilde{g}_3 = \frac{1}{16}\left(
    \frac{\alpha_1\alpha_2}{3}-\frac{2\alpha_1^3}{27}-\alpha_3
    \right),
\end{align}
\end{subequations}
given that $\alpha_1=C_2/C_3$, $\alpha_2=C_1/C_3$ and $\alpha_3=C_0/C_3$, and the respected coefficients defined as
\begin{subequations}\label{eq:C1C2C3}
\begin{align}
  & C_0 = r_D^4,\\
  & C_1 = 4 r_D^4,\\
  & C_2 = r_D^2\left(6r_D^2+\mA\right),\\
  & C_3 = r_D \left(4 r_D^3+2 \mA r_D+\mB\right).\label{eq;C3}
\end{align}
\end{subequations}
On the other hand, $r_F$ is the point of no return, from which, the photons will be captured bu the black hole ({deflection of the second kind (DSK)}).  Applying the same method we pursued to calculate the DFK, at the approaching point $r_F$, the DSK is found to obey the equation
\begin{equation}\label{eq:r(gamma)_cap}
    r_f(\gamma) = \frac{\left[1+u_f(\gamma) \right] r_F}{u_f(\gamma)},
\end{equation}
where
\begin{equation}\label{eq:u(gamma)_cap}
    u_f(\gamma) = 4\wp\left(-\frac{\omega_0\sqrt{\tilde{C_3}}\,\gamma}{M r_F}\right)-\frac{\tilde{\alpha}_1}{3},
\end{equation}
with the Weierstra$\ss$ coefficients having the same form of expressions as those in Eqs.~\eqref{eq:tg2tg3}, and  $\tilde{\alpha}_1=\tilde{C}_2/\tilde{C}_3$, $\tilde{\alpha}_2=\tilde{C}_1/\tilde{C}_3$ and $\tilde{\alpha}_3=\tilde{C}_0/\tilde{C}_3$, that relate respectively to the same coefficients in Eqs.~\eqref{eq:C1C2C3},  by doing the exchange $r_D\rightarrow r_F$. In Fig.~\ref{fig:rdrf}, the DFK and DSK have been plotted for the light rays approaching the black hole, for definite dynamical parameters. As it is expected, the DFK initiates from the turning point $r_D$ and escape to infinity, and can therefore, reach a distant observer. This concept, if treated for the trajectories with angular components, has the significance of bending of light in curved spacetimes.
{It is worth notifying the difference between the two types of the DFKs, as indicated in panel (a) of the figure. As it is expected, the more $\omega_0$ increases toward its critical value $\omega_U$, the more the trajectories are inclined towards the black hole. In order to find a switching value for $\omega_0$, at which the DFKs change their deflecting character, let us recall that $r_D\equiv r_D(\omega_0)$ and $r_F\equiv r_F(\omega_0)$, whose behaviors have been plotted in Fig.~\ref{fig:rDomega0}. As it is observed in the figure, there is a region of the steepest declination of $r_D$, located at the vicinity of $\omega_U$. This region starts in accordance with a switching value $\omega_0^\mathrm{e}$, from which, the trajectories start to change their deflecting character.
}
On the other hand, the DSK starts from $r_F$ and ends in falling onto the singularity. Hence, light rays that are engaged in this process, will never go beyond the distance $r_F$ from the black hole and cannot reach an observer at infinity. {Note that, in the case of $\omega_0=\omega_U$, the light rays will approach at the point $r_F<r_U<r_D$, that satisfy the equation $V'(r_U) = 0$. As mentioned above, this corresponds to the UCO. In Fig. \ref{fig:ru}, this kind of orbit has been plotted, regarding its first and the second kinds (UCOFK and UCOSK), by exploiting the equations of motion obtained above, applied for the radial distance $r_U$. The UCOFK, in particular, is responsible for the formation of the black hole shadow.}%

{\subsubsection{The capture zone}}

\begin{figure}[t]
\begin{center}
\includegraphics[width=6.0cm]{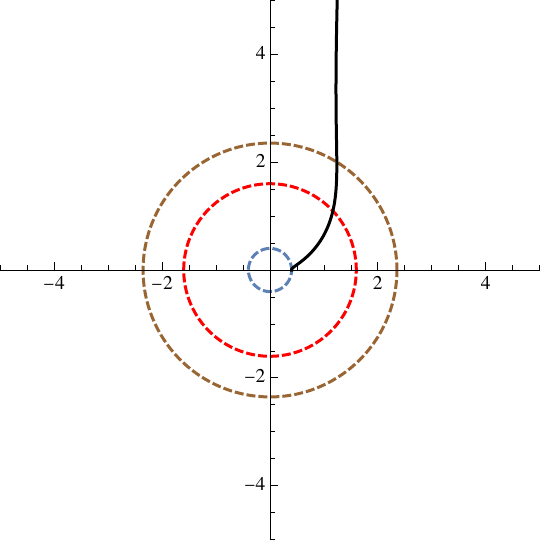}
\end{center}
\caption{{The radial capture. The outer circle corresponds to $r_U$.}}
\label{fig:rc}
\end{figure}

In addition to the photons with $\omega_0<\omega_U$ that approach the black hole from the radial distance $r_F$, those incident photons with $\omega_0>\omega_U$, also become completely unstable in the region dominant by the effective potential, and are captured by the black hole (see Fig. \ref{fig:EffectivePotential_typical}). The form of the equation of motion for such photons, is the same as those in Eqs. \eqref{eq:r(gamma)_def} and \eqref{eq:r(gamma)_cap}, but the point of approach can be any point $r_I>r_{+}$. In Fig. \ref{fig:rc}, an example of this kind of orbit has been plotted.


\subsection{The evolution of the polar angle (the $\theta$-motion) }

The $\theta$-motion is governed by  Eqs. \eqref{basiceqstheta} and \eqref{basiceqs2c}, for which, the condition \eqref{H62} implies
\begin{equation}\label{eq:Theta(theta)}
\Theta(\theta)=\mathscr{Q}-f_\theta-\cos^2\theta\left(L^2\csc^2\theta-a^2\omega_0^2\right)\geq 0.
\end{equation}
that can be recast as
\begin{equation}\label{eq:Theta(theta)2}
    \Theta(\theta) = a^2 \cos^2\theta\Big[\big(\omega_0-\sqrt{W(\theta)}\,\big)\,\big(\omega_0+\sqrt{W(\theta)}\,\big) \Big],
\end{equation}
where
\begin{equation}\label{Wtheta}
W(\theta)=\left(\frac{L \csc \theta }{a}\right)^2-\left(\mathscr{Q}-f_\theta\right) \left(\frac{\sec \theta }{a}\right)^2,
\end{equation}
is the angular gravitational potential felt by the light rays in the plasma. 
{Essentially, this potential has the general form as defined, for example, for the case of null geodesics in Ref.~\cite{Schee:2009}. This general form can be recovered by letting $f_\theta=0$ in Eq.~\eqref{Wtheta}.}
We define
\begin{equation}
\tilde{\eta}= \eta-{f_\theta\over \omega_0^2}\equiv \frac{\mathscr{Q}-f_\theta}{\omega_0^2},
\end{equation}
for more convenience. Now, performing the change of variable $z=\cos\theta$, Eq. \eqref{basiceqstheta} can be recast as
\begin{equation}\label{thetazeta}
-{M\over \omega_0}\frac{\ed z}{\ed \gamma}=\sqrt{\Theta_{z}},
\end{equation}
where
\begin{equation}\label{thetazeta2}
\Theta_{z}=\tilde{\eta}-(\tilde{\eta}+\xi^2-a^2)z^2-a^{2}z^4,
\end{equation}
and the condition $\Theta_z>0$ is required. Clearly, the characteristics of the motion depend directly on the nature of $\Theta_z$. Accordingly, we assess the equation of motion \eqref{thetazeta}, separately, for the cases $\tilde{\eta}>0$, $\tilde{\eta}=0$, and $\tilde{\eta}<0$. These cases completely categorize the types of the $\theta$-motion.

\subsubsection{The case of $\tilde{\eta}>0$}

\begin{figure}[t]
	\begin{center}
		\includegraphics[width=6cm]{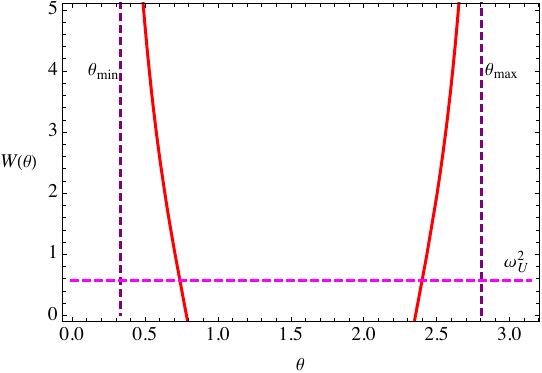}
		\includegraphics[width=6cm]{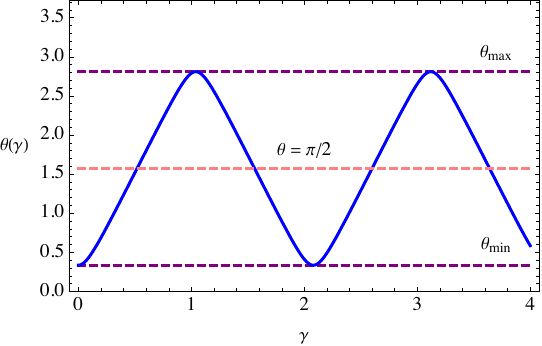}
	\end{center}
	\caption{{The angular effective potential and the temporal evolution of the coordinate $\theta$ for the case of $\tilde{\eta}>0$, plotted for  $\mathscr{Q} = 9M^2$ and $f_\theta = 1 M^2$. To plot $\theta(\gamma)$, we have considered ${\omega_0 = 0.755}$.}}
	\label{fig:W1}
\end{figure}

\begin{figure}[h]
\begin{center}
\includegraphics[width=5.5cm]{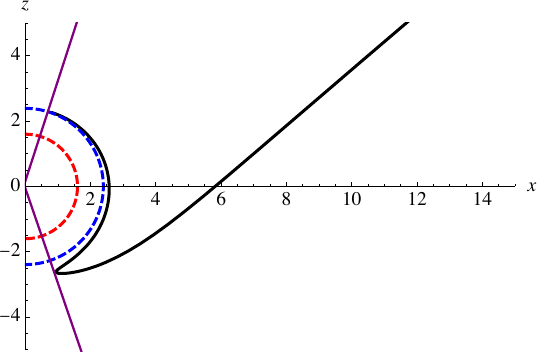} (a)
\includegraphics[width=5.5cm]{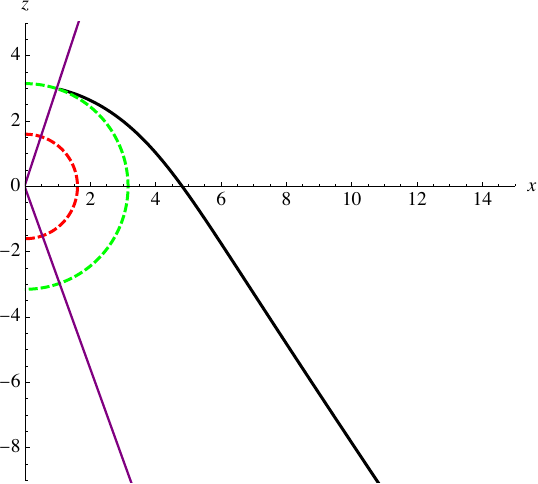} (b)
\includegraphics[width=3cm]{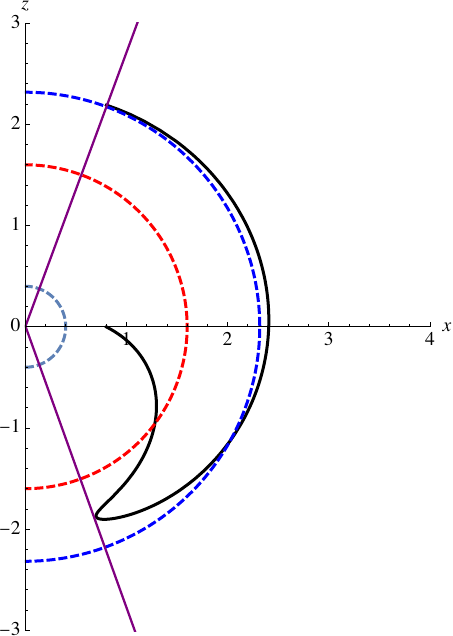} (c)
\includegraphics[width=3cm]{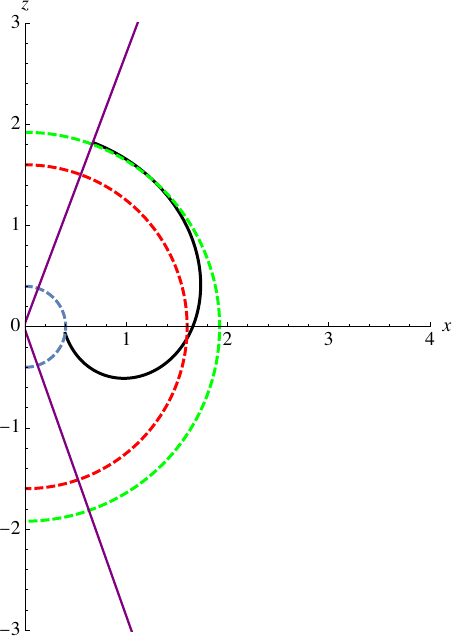} (d)
\includegraphics[width=3cm]{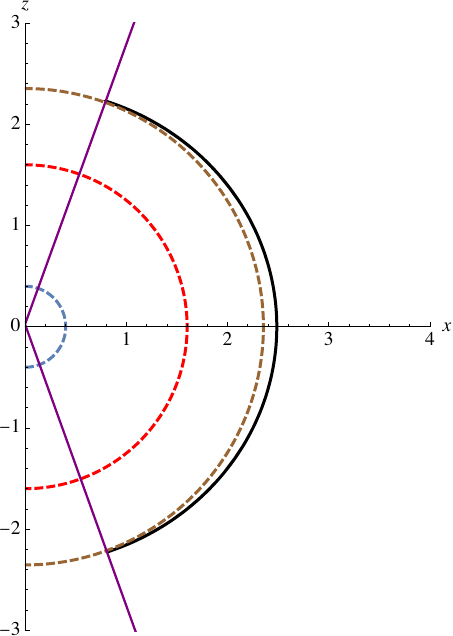} (e)
\includegraphics[width=3cm]{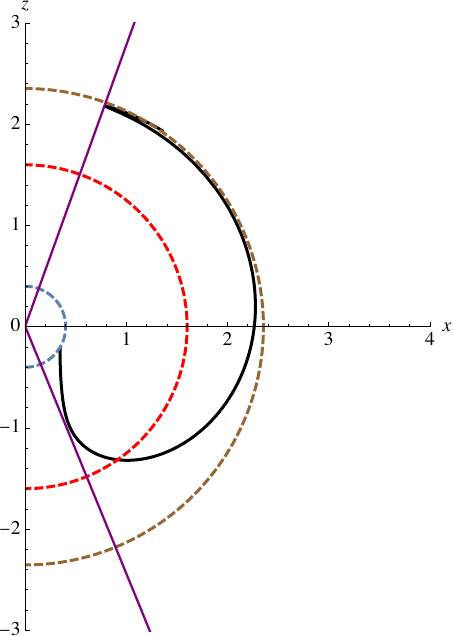} (f)
\includegraphics[width=3cm]{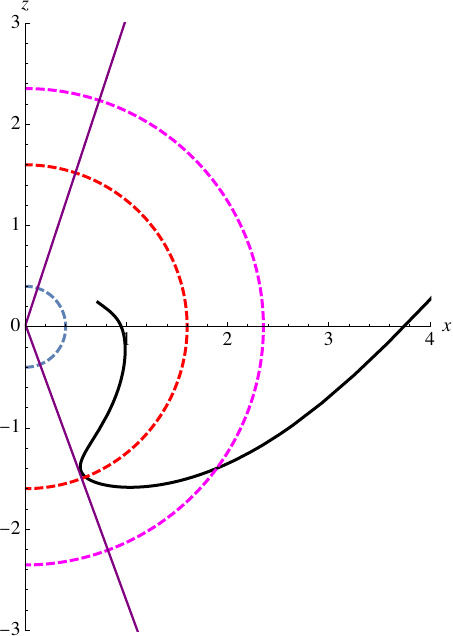} (g)
\end{center}
\caption{{The cross-sectional behaviors of the deflecting, critical, and  capturing trajectories, for the case of $\tilde{\eta}>0$, inside the cone of confinement, in the $z$-$x$ plane. The diagrams indicate (a) DFK for ${\omega_0 = 0.755}$, (b) DFK for $\omega_0 = 0.70$, (c) DSK for ${\omega_0 = 0.755}$, (d) DSK for $\omega_0 = 0.70$, (e) UCOFK, (f) UCOSK, and (g) radial capture for $\omega_0 = 1$.
}}
\label{fig:2D}
\end{figure}

In this case, the expression in Eq.~\eqref{thetazeta2} remains unchanged, and the condition $\Theta_{z}>0$, confines the $z$ parameter in the domain  $-z_{\min}\leq z \leq z_{\max}$, where
\begin{subequations}
\begin{align}
 &   z_{\max}^2={\chi_0\over2\, a^2}\left(\sqrt{1+{4a^2\,\tilde{\eta}\over \chi_0^2}}-1\right),\label{zmax}\\
&   z_{\min}=-z_{\max},
\end{align}
\end{subequations}
in which, $\chi_0=\xi^2+\tilde{\eta}-a^2>0$. The mentioned domain, corresponds to the polar range $\theta_{\min}\leq \theta \leq \theta_{\max}$, given that  $\theta_{\min}=\arccos\left(z_{\max}\right)$ and $\theta_{\max}=\arccos\left(-z_{\max}\right)$. This range defines a cone, to which, the test particles' motion is confined. Having this in mind, we can solve the equation of motion \eqref{thetazeta} by direct integration, resulting in 
\begin{equation}\label{h.3}
\theta(\gamma)=
\arccos\left(z_{\max} -\frac{3}{12\wp\left(\kappa_0 \gamma\right)+\psi_0}\right),
\end{equation}
where
\begin{subequations}
\begin{align}
    & \kappa_0 = \frac{\omega_0 a \sqrt{2 z_{\max}\left(z_0^2+z_{\max}^2\right)}}{M},\label{eq:kappa0}\\
    & \psi_0 = \frac{z_0^2+5 z_{\max}^2}{2z_{\max}\left(z_0^2+z_{\max}^2\right)},\label{eq:psi0}
\end{align}
\end{subequations}
with 
\begin{equation}\label{z0}
    z_{0}^2={\chi_0\over2\, a^2}\left(\sqrt{1+{4a^2\,\tilde{\eta}\over \chi_0^2}}+1\right),
\end{equation}
and 
\begin{subequations}\label{eq:g2,3}
\begin{align}
 &   g_2 = \frac{z_0^4+z_{\max}^4-14 z_0^2 z_{\max}^2}{48 z_{\max}^2 \left(z_0^2+z_{\max}^2\right)^2},\label{g2}\\
& g_3=\frac{33 z_0^4 z_{\max}^2-33 z_0^2 z_{\max}^4+z_0^6-z_{\max}^6}{1728 z_{\max}^3 \left(z_0^2+z_{\max}^2\right)^3},\label{g3} 
\end{align}
\end{subequations}
are the respected Weierstra$\ss$ invariants. Using the analytical solution \eqref{h.3}, the temporal evolution of $\theta(\gamma)$ for the case of $\tilde{\eta}>0$ has been shown in Fig. \ref{fig:W1}, together with the behavior of $W(\theta)$. As it can be observed from the behavior of $W(\theta)$, the values of energy that rely in the region $\omega_0\leq\omega_U$ are allowed. The light rays, therefore, can opt all kinds of possible orbits that were discussed in the previous section. Accordingly, and applying the analytical solutions of the radial coordinate, the cross-sectional behaviors of the above orbits have been plotted in Fig. \ref{fig:2D}, for the case of $\tilde{\eta} >0$, inside the cone of confinement in the $z$-$x$ plane (i.e. for $\phi = 0$).

\subsubsection{The case of $\tilde{\eta}=0$}

\begin{figure}[t]
	\begin{center}
		\includegraphics[width=6cm]{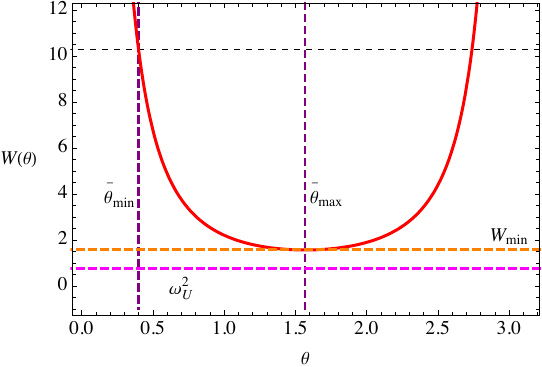}
		\includegraphics[width=6cm]{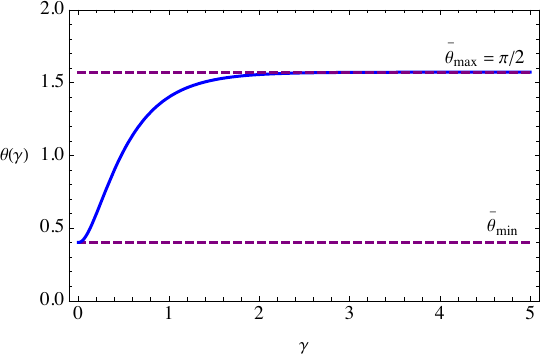}
	\end{center}
	\caption{{The angular effective potential and the temporal evolution of the coordinate $\theta$, for the case of $\tilde{\eta}=0$ (which here corresponds to $\mathscr{Q} = f_\theta = 9 M^2$). To plot $\theta(\gamma)$, we have considered $\omega_0 = \sqrt{10.30}$.}}
	\label{fig:W2}
\end{figure}

\begin{figure}[t]
	\begin{center}
		\includegraphics[width=6cm]{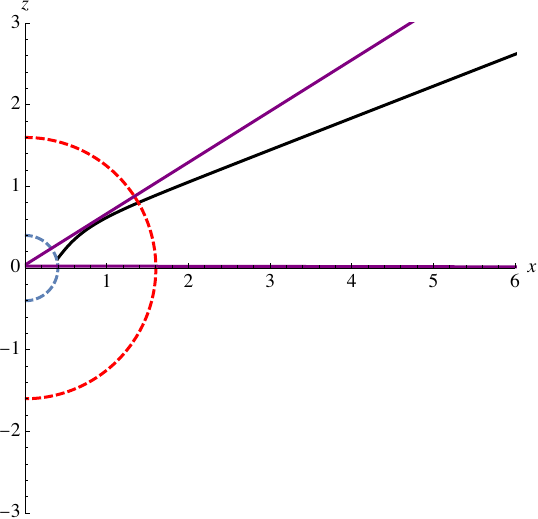}
	\end{center}
	\caption{{The capturing trajectory for the case of  $\tilde{\eta}=0$, inside the cone of confinement, plotted for $\omega_0 = \sqrt{10.30}$. The outer and inner circles indicate, respectively, $r_+$ and $r_-$.}}
	\label{fig:2D0}
\end{figure}

The parameter $z$, in this case, is confined to the domain $\bar{z}_{\min}\leq z\leq\bar{z}_{\max}$ with $\bar{z}_{\min} = 0$ and $\bar{z}_{\max} = \sqrt{1-(\xi/a)^2}$. This domain corresponds to the cone $\bar{\theta}_{\min}\leq\theta\leq\pi/2$, where $\bar{\theta}_{\min}=\arccos\left(\bar{z}_{\max}\right)$. In this case, the effective angular potential \eqref{Wtheta} takes the form
\begin{equation}\label{Wtheta0}
   W(\theta) = \left(\frac{L \csc\theta}{a}\right)^2.
\end{equation}
It is straightforward to see that $W'(\theta) = 0$ gives $\bar{\theta}_{0}=\pi/2=\bar{\theta}_{\max}$, according to which, the minimum of the angular effective potential, $W_{\min} = W(\pi/2)$, is obtained. One can therefore infer that the case of $\tilde{\eta} = 0$ also allows for a stable polar equatorial motion. Taking into account $\theta(0) = \bar{\theta}_{\min}$, the direct integration of Eq.~\eqref{thetazeta} yields
\begin{equation}\label{h.4}
\theta(\gamma)=
\arccos\left(\sqrt{1-\left(\frac{\xi}{a}\right)^2}\,\sech\left( \kappa_1\gamma\right)\right),
\end{equation}
which implies $a^2>\xi^2$, and we have defined $ \kappa_1=\omega_0\sqrt{a^2-\xi^2}/M$. Accordingly, the temporal evolution of $\theta(\gamma)$ and the corresponding angular effective potential have been plotted in Fig. \ref{fig:W2}. As it can be observed, in contrast with the case of $\tilde{\eta}>0$, the allowed energies for the case of $\tilde{\eta} = 0$ are higher than their critical value (i.e. $W_{\min} > \omega_U^2$), and therefore, only the capturing trajectories are possible, which in Fig. \ref{fig:2D0}, has been demonstrated inside the cone of confinement.


\subsubsection{The case of $\tilde{\eta}<0$}

\begin{figure}[t]
	\begin{center}
		\includegraphics[width=6 cm]{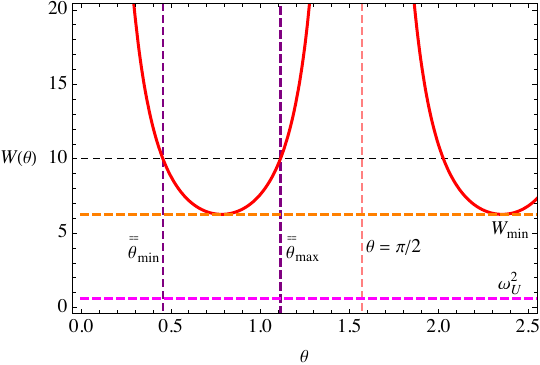}
		\includegraphics[width=6 cm]{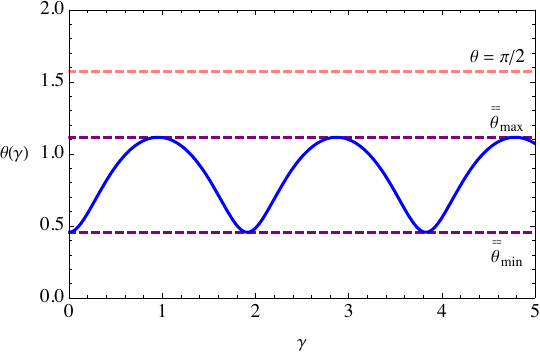}
	\end{center}
	\caption{The angular effective potential and the temporal evolution of the coordinate $\theta$ for the case of $\tilde{\eta}<0$, plotted for $\mathscr{Q} = 9M^2$ and $f_\theta = 10 M^2$. To plot $\theta(\gamma)$, we have considered $\omega_0 = \sqrt{10}$.}
	\label{fig:W3}
\end{figure}

\begin{figure}[t]
	\begin{center}
		\includegraphics[width=4 cm]{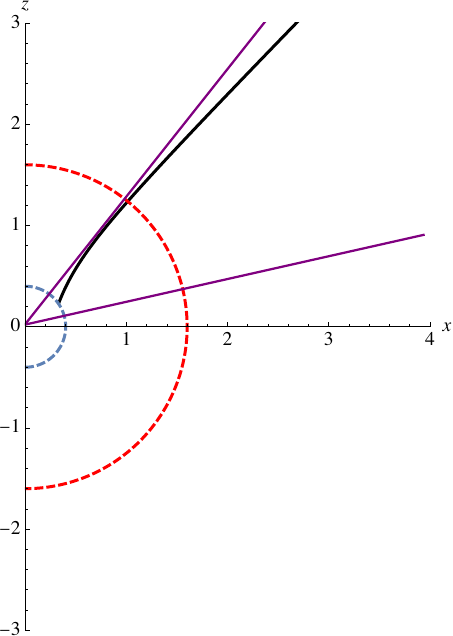}
	\end{center}
	\caption{The capturing trajectory for the case of $\tilde{\eta}<0$, plotted for $\omega_0=\sqrt{10}$.}
	\label{fig:2Dn}
\end{figure}

Under this condition, the expression in Eq.~\eqref{thetazeta2} takes the form
\begin{equation}
\Theta_{z}=-| \tilde{\eta}|+ (| \tilde{\eta}|+a^2-\xi^2)z^2-a^{2}z^4,
\label{h.1}
\end{equation}
and $\Theta_{z}>0$ requires $| \tilde{\eta}|+a^2-\xi^2>0 $, which is satisfied inside the domain $\bar{\bar{z}}_{\min}\leq z \leq \bar{\bar{z}}_{\max}$, where 
\begin{subequations}
\begin{align}
  & \bar{\bar{z}}_{\min} = \mu_0\sin\left( {1\over 2}\arcsin\left(\mu_1\right) \right), \label{z1} \\
  & \bar{\bar{z}}_{\max} = \mu_0\cos\left( {1\over 2}\arcsin\left(\mu_1 \right) \right),\label{z2}
\end{align}
\end{subequations}
with
\begin{subequations}
\begin{align}
  & \mu_0 = {\sqrt{ | \tilde{\eta}|+a^2-\xi^2}\over a},\label{zeta1}\\
  & \mu_1 = {2a \sqrt{| \tilde{\eta}|}\over | \tilde{\eta}|+a^2-\xi^2}.\label{zeta2} 
\end{align}
\end{subequations}
The corresponding particle-cone is therefore confined to $\bar{\bar{\theta}}_{\min}\leq\theta\leq \bar{\bar{\theta}}_{\max}$, where $\bar{\bar{\theta}}_{\min}=\arccos\left(\bar{\bar{z}}_{\max}\right)$ and $\bar{\bar{\theta}}_{\max}=\arccos\left(\bar{\bar{z}}_{\min}\right)$, and the respected effective potential is
\begin{equation}\label{Wtheta0}
   W(\theta)=\left(\frac{L \csc \theta }{a}\right)^2+\left|\mathscr{Q}-f_\theta\right| \left(\frac{\sec \theta }{a}\right)^2.
\end{equation}
Once again, to determine the possible stable polar orbits we solve $W'(\theta) = 0$, which yields
\begin{equation}\label{eq:plormin}
    \bar{\bar{\theta}}_{0} = \arctan\left(|\tilde{\eta}|^{\frac{1}{4}}\sqrt{\frac{\omega_0 \left(L+\omega_0 |\tilde{\eta}|^\frac{1}{2}\right)}{L^2+\omega_0^2 |\tilde{\eta}|}}\right).
\end{equation}
This value satisfies $\bar{\bar{\theta}}_{\min}\leq\bar{\bar{\theta}}_{0}\leq \bar{\bar{\theta}}_{\max}$, and corresponds to the minimum of the angular effective potential for the case of $\tilde{\eta}<0$. To find the analytical solution for $\theta(\gamma)$, we pursue the same method as in the case of $\tilde{\eta}>0$, that provides
\begin{equation}\label{h.5}
\theta(\gamma)=
\arccos\left(\bar{\bar{z}}_{\max} -\frac{3}{12\wp\left(\kappa_2 \gamma\right)+\varphi_0}\right),
\end{equation}
where 
\begin{subequations}
\begin{align}
    & \kappa_2 = \frac{\omega_0 a \sqrt{2 \bar{\bar{z}}_{\max}\left(2\bar{\bar{z}}_{\max}^2-\mu_0^2\right)}}{M},\label{eq:kappa2}\\
    & \varphi_0 = \frac{6\bar{\bar{z}}_{\max}^2-\mu_0^2}{2 \bar{\bar{z}}_{\max}\left(2\bar{\bar{z}}_{\max}^2-\mu_0^2\right)}\label{eq:varphi0},
\end{align}
\end{subequations}
and the corresponding Weierstra$\ss$ invariants are
\begin{subequations}\label{eq:gb}
\begin{align}
    & \bar{\bar{g}}_2 = \frac{\mu_0^4+12 \mu_0^2 \bar{\bar{z}}_{\max}^2-12 \bar{\bar{z}}_{\max}^4}{48 \left(\mu_0^2 \bar{\bar{z}}_{\max}-2 \bar{\bar{z}}_{\max}^3\right)^2},\label{gb2}\\
    & \bar{\bar{g}}_3 = -\frac{\mu_0^2 \left(\mu_0^4-36 \mu_0^2 \bar{\bar{z}}_{\max}^2+36 \bar{\bar{z}}_{\max}^4\right)}{1728 \left(2 \bar{\bar{z}}_{\max}^3-\mu_0^2 \bar{\bar{z}}_{\max}\right)^3}.\label{gb3}
\end{align}
\end{subequations}
As in the previous cases, we have plotted the respected angular effective potential and the temporal evolution of the $\theta$-coordinate, in Fig. \ref{fig:W3}, for the case of $\tilde{\eta}<0$, which similar to the case of $\tilde{\eta} = 0$, indicates that $W_{\min}>\omega_U^2$. Hence, only the capturing trajectories are allowed (see Fig. \ref{fig:2Dn}).                        


\subsection{The evolution of the azimuth angle (the $\phi$-motion) }

\begin{figure}[t]
\begin{center}
\includegraphics[width=4.1cm]{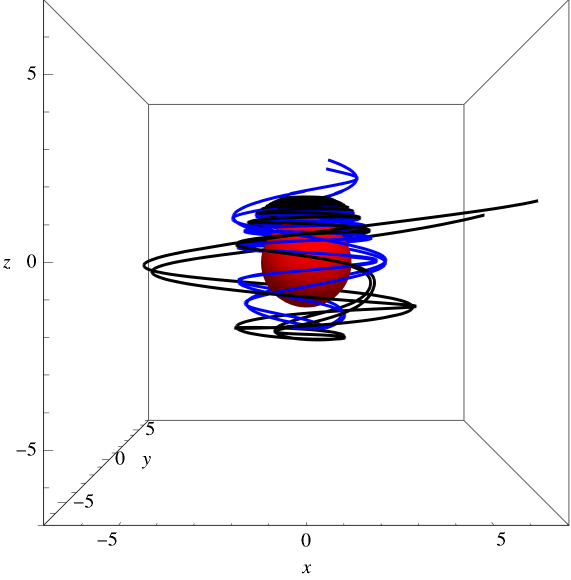} (a)
\includegraphics[width=4.1cm]{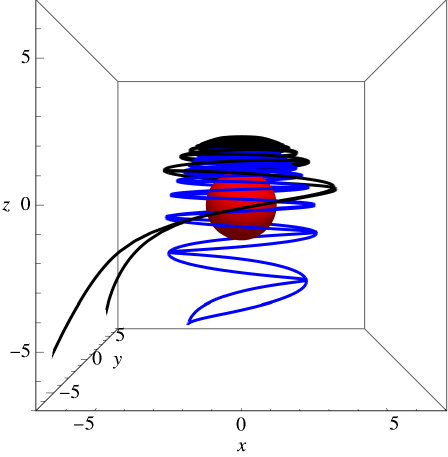} (b)
\includegraphics[width=4.1cm]{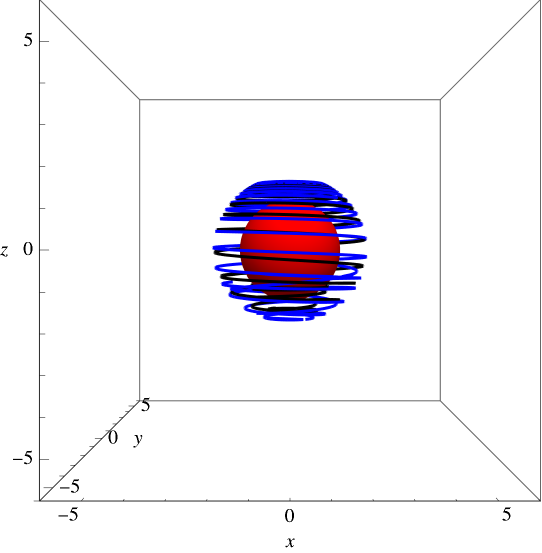} (c)
\includegraphics[width=4.1cm]{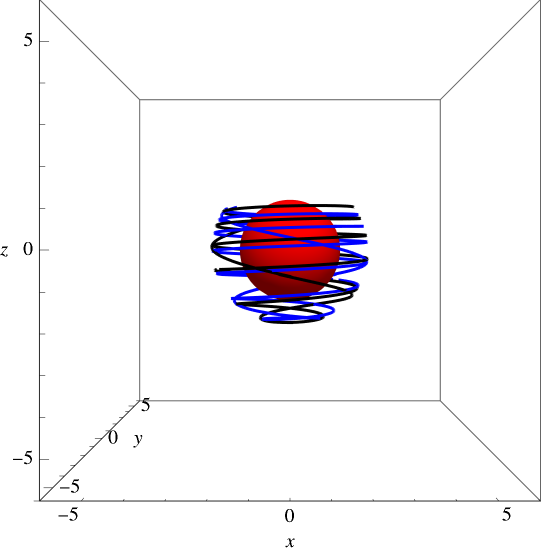} (d)
\includegraphics[width=4.1cm]{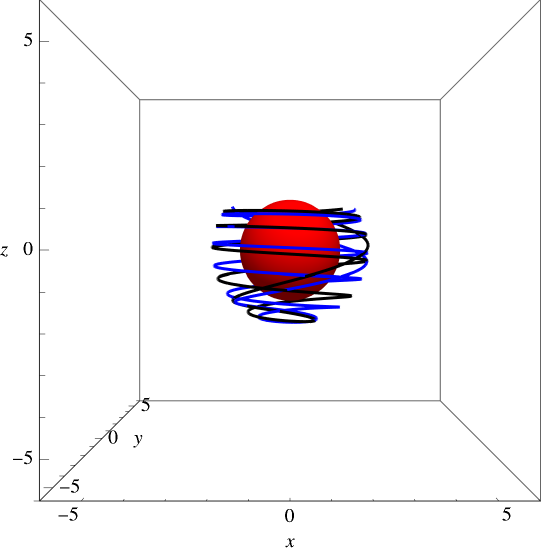} (e)
\includegraphics[width=4.1cm]{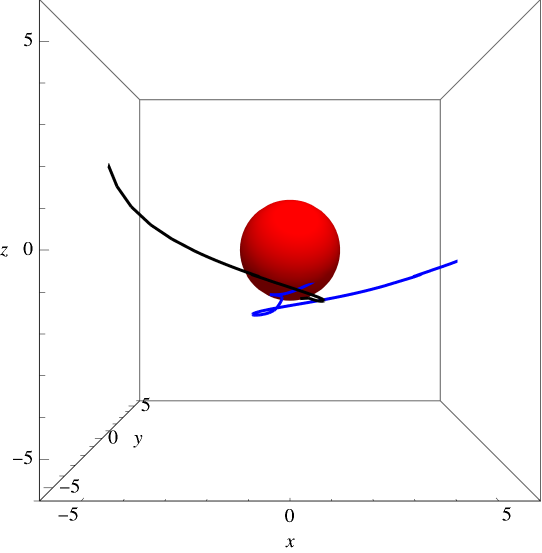} (f)
\includegraphics[width=4.15cm]{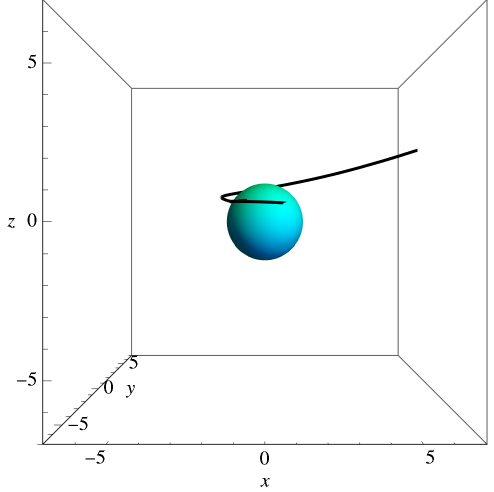} (g)
\includegraphics[width=4.1cm]{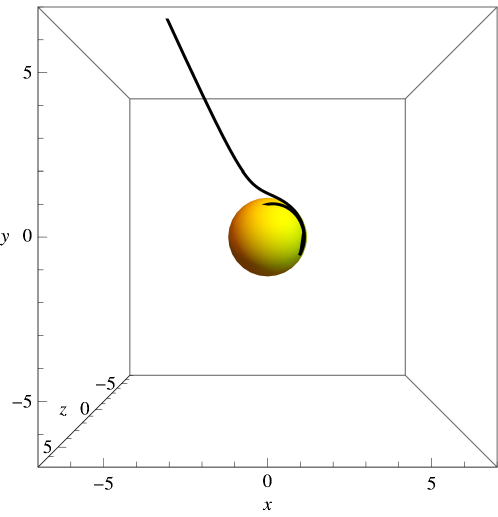} (h)
\end{center}
\caption{{The three-dimensional simulations of the possible orbits plotted for $\mathscr{Q} = 9M^2$, corresponding to the three cases of $\tilde{\eta}>0$ (a-f) where the black trajectories have been calculated for $f_r=f_\theta=1M^2$ and the blue ones indicate the null trajectories in the vacuum Kerr spacetime (i.e. for $f_r=f_\theta=0$), $\tilde{\eta}=0$ (g), and $\tilde{\eta}<0$ (h). The sphere in the middle indicates the event horizon. The diagrams correspond to (a) DFK for ${\omega_0 = 0.755}$, (b) DFK for $\omega_0 = 0.70$, (c) DSK for ${\omega_0 = 0.755}$, (d) UCOFK, (e) UCOSK, (f) capture for $\omega_0 = 1$, (g) capture for $\omega_0 = \sqrt{10.30}$, (h) capture for $\omega_0 = \sqrt{10}$. Theses last two trajectories do not have a vacuum counterpart, for the above chosen Carter's constant.
}}
\label{fig:3D}
\end{figure}

A feasible approach to calculate the $\phi$-motion, is to divide the integral equation \eqref{basiceqsphi} into $\theta$-dependent and $r$-dependent parts, in a way that we get 
\begin{equation}\label{eq:h.4}
    \phi(\gamma) =  \Phi_\theta(\gamma)+ \Phi_r(\gamma),
\end{equation}
with 
\begin{subequations}\label{h.4}
\begin{align}
 &  \Phi_\theta(\gamma) = \int^{\theta(\gamma)}_{\theta_{\min}}
\frac{L \ed\theta}{\sin^{2}\theta\sqrt{\Theta(\theta)}}, \label{h.4a}\\
 &   \Phi_r(\gamma) = \int^{r(\gamma)}_{r_{i}}
\frac{\left(a^2L+2Ma\omega_0 r\right)\ed r}{\Delta\sqrt{\mathcal{R}(r)}}, \label{h.4b}
\end{align}
\end{subequations}
in which, the minimum value of the $\theta$ coordinate has been assumed to coincide with the initial point $r_i$, which can be set as either of the turning points. According to the fact that the general form of the equation of motion has been considered, this assumption does not affect the final analytical results for the $\phi$-motion. Direct integration of the integral \eqref{h.4a} results in the following cases: 
\begin{itemize}
\item{{For $\tilde{\eta}>0$ we get 
\begin{equation}\label{eq:h.42}
\Phi_{\theta}(\gamma) = \mathcal{K}_0 \big[\mathcal{K}_1 \mathcal{F}_1(U_\theta) - \mathcal{K}_2 \mathcal{F}_2(U_\theta)-\kappa_0 \gamma\big],
\end{equation}
in which, $\kappa_0$ has been given in Eq. \eqref{eq:kappa0}, and (with $j=1,2$)
\begin{equation}\label{eq:h.43}
\mathcal{F}_j(U_\theta) = \frac{1}{\wp'(\upsilon_j)}
\left[\ln\left(
\frac{\sigma\big(\ss(U_\theta)-\upsilon_j\big)}
{\sigma\big(\ss(U_\theta)+\upsilon_j\big)}\right)
+ 2 \ss(U_\theta) \zeta(\upsilon_j)
\right],
\end{equation}
where, $\wp'(\upsilon) \equiv \frac{\ed}{\ed\upsilon}\wp(\upsilon;g_2,g_3)$, which here, is given in terms of the corresponding Weierstra$\ss$ invariants $g_2$ and $g_3$, as in Eqs. \eqref{eq:g2,3}. Furthermore,
\begin{subequations}\label{eq:g1Utheta}
\begin{align}
   &   \upsilon_1 = \ss\left(-\frac{\psi_0}{12}-\frac{1}{4 [1-z_{\max}]}\right),\\
   &   \upsilon_2 = \ss\left(-\frac{\psi_0}{12}+\frac{1}{4 [1+z_{\max}]}\right),\\
   &  U_\theta=\frac{1}{4 (z_{\max}-\cos\theta)}-\frac{\psi_0}{3},
\end{align}
\end{subequations}
in which, $\ss(x)\equiv\wp^{-1}\left(x;g_2,g_3\right)$ is the inverse Weierstra$\ss$ian $\wp$ function, $z_{\max}$ and $\psi_0$ are given in Eqs. \eqref{zmax} and \eqref{eq:psi0}, and
\begin{subequations}\label{eq:K1K2}
\begin{align}
& \mathcal{K}_0 = \frac{\xi}{a \sqrt{2z_{\max} \left(z_{\max}^2+z_0^2\right)}\, (1-z_{\max}) (1+z_{\max})},\\
    & \mathcal{K}_1 = \frac{1+z_{\max}}{8 (1-z_{\max})},\\
    & \mathcal{K}_2 = \frac{1-z_{\max}}{8 (1+z_{\max})}.
\end{align}
\end{subequations}
The function in Eq.~\eqref{eq:h.43}, also depends on the Weierstra$\ss$ian Zeta and Sigma functions (see appendix \ref{app:A}). Note that, for the sake of simplicity in the demonstration of the trajectories, we will set $\phi(\theta_{\min})=0$ as the initial condition. The complete $\gamma$-dependent expression for $\Phi_\theta(\gamma)$ is then obtained by substituting $\theta\rightarrow \theta(\gamma)$ in the above relations.}}
\item{{For $\tilde{\eta} = 0$, pursuing the same mathematical methods, we find
\begin{equation}\label{eq:h.44}
    \Phi_{\theta}(\gamma) = \bar{\mathcal{K}}_0 \big[\bar{\mathcal{K}}_1 \bar{\mathcal{F}}_1(\bar{U}_\theta) - \bar{\mathcal{K}}_2 \bar{\mathcal{F}}_2(\bar{U}_\theta)-
    \ss\left(\bar{U}_\theta\right)\big],
\end{equation}
in which, $\bar{\mathcal{F}}_\theta(\theta)$ has the same expression as in Eq.~\eqref{eq:h.43}, considering the exchanges
\begin{subequations}\label{eq:g1barUtheta}
\begin{align}
   &   \upsilon_1\rightarrow \bar{\upsilon}_1 =  \ss\left(-\frac{5}{24 \bar{z}_{\max}}-\frac{1}{4 [1-\bar{z}_{\max}]}\right),\\
   &   \upsilon_2\rightarrow \bar{\upsilon}_2 =  \ss\left(-\frac{5}{24 \bar{z}_{\max}}+\frac{1}{4 [1+\bar{z}_{\max}]}\right),\\
   &  U_\theta\rightarrow\bar{U}_\theta=\frac{1}{4 (\bar{z}_{\max}-\cos\theta)}-\frac{5}{24 \bar{z}_{\max}},
\end{align}
\end{subequations}
and the corresponding Weierstra$\ss$ invariants are
\begin{subequations}\label{eq:g2,3-eta0}
\begin{align}
 & \bar{g}_2 = \frac{1}{48 \bar{z}_{\max}^2},\\
& \bar{g}_3=-\frac{1}{1728 \bar{z}_{\max}^3}.
\end{align}
\end{subequations}
Furthermore,
\begin{subequations}\label{eq:Kb1Kb2}
\begin{align}
& \bar{\mathcal{K}}_0 = \frac{\xi}{a \sqrt{2\bar{z}_{\max}^3}\, (1-\bar{z}_{\max}) (1+\bar{z}_{\max})},\\
    & \bar{\mathcal{K}}_1 = \frac{1+\bar{z}_{\max}}{8 (1-\bar{z}_{\max})},\\
    & \bar{\mathcal{K}}_2 = \frac{1-\bar{z}_{\max}}{8 (1+\bar{z}_{\max})}.
\end{align}
\end{subequations}
}
\item{{For $\tilde{\eta}<0$ we obtain
\begin{equation}\label{eq:h.45}
    \Phi_{\theta}(\gamma) = \bar{\bar{\mathcal{K}}}_0 \big[\bar{\bar{\mathcal{K}}}_1 \bar{\bar{\mathcal{F}}}_1(\bar{\bar{U}}_\theta) - \bar{\bar{\mathcal{K}}}_2 \bar{\bar{\mathcal{F}}}_2(\bar{\bar{U}}_\theta)-\kappa_2 \gamma\big],
\end{equation}
with $\kappa_2$ given in Eq. \eqref{eq:kappa2}. To obtain $\bar{\bar{\mathcal{F}}}_j(\bar{\bar{U}}_\theta)$, we need to apply
\begin{subequations}\label{eq:g1barbarUtheta}
\begin{align}
   &   \upsilon_1\rightarrow\bar{\bar{\upsilon}}_1  =\ss\left(-\frac{\varphi_0}{12}-\frac{1}{4[1-\bar{\bar{z}}_{\max}]}\right),\\
   &   \upsilon_2\rightarrow\bar{\bar{\upsilon}}_1  =\ss\left(-\frac{\varphi_0}{12}+\frac{1}{4[1+\bar{\bar{z}}_{\max}]}\right),\\
   &  U_\theta\rightarrow\bar{\bar{U}}_\theta=\frac{1}{4(\bar{\bar{z}}_{\max}-\cos\theta)}-\frac{\varphi_0}{3},
\end{align}
\end{subequations}
in Eqs.~\eqref{eq:h.43} and \eqref{eq:g1Utheta}, with $\bar{\bar{z}}_{\max}$, $\mu_0$ and $\varphi_0$, defined, respectively, in Eqs. \eqref{z2}, \eqref{zeta1} and \eqref{eq:varphi0}. The corresponding Weierstra$\ss$ invariants are the same as $\bar{\bar{g}}_2$ and $\bar{\bar{g}}_3$, given in Eqs. \eqref{eq:gb}, and
\begin{subequations}\label{eq:Kb1Kb2}
\begin{align}
& \bar{\bar{\mathcal{K}}}_0 = \frac{\xi}{a \sqrt{2\bar{\bar{z}}_{\max} \left(2\bar{\bar{z}}_{\max}^2-\mu_0^2\right)}\, (1-\bar{\bar{z}}_{\max}) (1+\bar{\bar{z}}_{\max})},\\
    & \bar{\bar{\mathcal{K}}}_1 = \frac{1+\bar{\bar{z}}_{\max}}{8 (1-\bar{\bar{z}}_{\max})},\\
    & \bar{\bar{\mathcal{K}}}_2 = \frac{1-\bar{\bar{z}}_{\max}}{8 (1+\bar{\bar{z}}_{\max})}.
\end{align}
\end{subequations}
}}}
\end{itemize}
The $r$-dependent integral \eqref{h.4b} provides 
\begin{equation}\label{eq:Phir}
    \Phi_r(\gamma) = \mathcal{K}_+ \mathcal{F}_+(U_r) - \mathcal{K}_- \mathcal{F}_-(U_r) - \tilde{B}\,\ss(U_r),
\end{equation}
in which
\begin{equation}\label{eq:Fpm}
   \mathcal{F}_\pm(U_r) = \frac{1}{\wp'(\varUpsilon_\pm)}\left[\ln\left(
\frac{\sigma\big(\varUpsilon_\pm-\ss(U_r)\big)}{\sigma\big(\varUpsilon_\pm+\ss(U_r)\big)}\right)
+ 2 \ss(U_r) \zeta(\varUpsilon_\pm)
\right],
\end{equation}
given that
\begin{subequations}\label{eq:Ur}
\begin{align}
&  \mathcal{K}_\pm = \frac{ r_i \left(a^2 L+2 M a \omega_0 r_i \right)+2 M a \omega_0 r_i (r_i-r_\pm)}{4 \omega_0\sqrt{\tilde{\alpha}} (r_i-r_\pm)^2(r_+-r_-)},\\
&  \tilde{B} = \frac{\left(a^2 L + 2 M a \omega_0 r_i\right)}{\omega_0\sqrt{\tilde{\alpha}}(r_i-r_+)(r_i-r_-)},\\
&  U_r\equiv U_r(\gamma) = \frac{\alpha_1}{12} -\frac{  {r_i}}{4\left[r(\gamma)-{r_i}\right]},\label{eq:Ur,beta}\\
& \varUpsilon_\pm = \ss\left(\frac{\alpha_1}{12} - \frac{r_i}{4\left[r_i-r_\pm\right]}\right),\label{eq:varUpsilonpm}
\end{align}
\end{subequations}
where the coefficient $\alpha_1$ and the Weierstra$\ss$ invariants, are the same as those given in Eqs. \eqref{eq:tg2tg3}, considering $r_D\rightarrow r_i$. Furthermore
\begin{equation}\label{eq:alphatbetat}
 \tilde\alpha = 2 \mathcal{A}+4 r_i^2+\frac{\mathcal{B}}{r_i}.
\end{equation}
Now, having in hand the analytical expressions for all the spatial coordinates, we are able to simulate the possible orbits, based on the simultaneous evolution of these coordinates. Respecting the allowed values of $\omega_0 $, in Fig. \ref{fig:3D}, the orbits that we have previously illustrated in Figs. \ref{fig:2D}, \ref{fig:2D0}, and \ref{fig:2Dn}, are demonstrated in the three-dimensional form, by applying the following Cartesian correspondents of the Boyer-Lindquist coordinates
\cite{Boyer:1967} 
\begin{subequations}\label{eq:Cartesian}
\begin{align}
  &  x(\gamma) = \sqrt{r^2(\gamma)+a^2}\sin\theta(\gamma)\cos\phi(\gamma),\\
  &  y(\gamma) = \sqrt{r^2(\gamma)+a^2}\sin\theta(\gamma)\sin\phi(\gamma),\\
  &  z(\gamma) = r(\gamma)\cos\theta(\gamma).
\end{align}
\end{subequations}

In presenting the figures, we have also considered the case of the vacuum Kerr spacetime, for which, the characteristic functions $f_r(r)$ and $f_\theta(\theta)$ in Eq.~\eqref{plasma2}, vanish identically. In fact, in the three-dimensional treatment of the trajectories, a fixed positive Carter's constant $\mathscr{Q}$, does not allow for the construction of the cases $\tilde{\eta}=0$ and $\tilde{\eta}<0$ in the vacuum Kerr spacetime. Hence, the possible comparison between the light propagation in the plasmic and vacuum Kerr spacetimes can only be done in the context of $\tilde{\eta}>0$ which corresponds to the first six diagrams of Fig.~\ref{fig:3D}. As indicated in these diagrams, the sensible differences between these media, are seen in the DFKs. While the light ray trajectories in plasma (black curves) occupy a wider spatial range  around the black hole, those in the vacuum (blue curves) are more close to the event horizon. Passing the plasma, will therefore, change the amount of light deflection and this can be inspected through the process of gravitational lensing (see below). The other types of trajectories do not show any sensible differences. Hence, in what follows we continue with the equatorial lens equation for the black hole.

\subsubsection{Gravitational lensing}\label{subsubsec:lensing}

Let us, for now, confine ourselves to the equatorial plan (with $\theta=\pi/2$), on which we have $\mathscr{Q} = f_\theta$. Therefore, by means of Eqs.~\eqref{basiceqsR} and \eqref{basiceqsphi}, the differential equation that governs the gravitational lensing is 
\begin{equation}\label{eq:lensingDiff}
    \left.\frac{\ed\phi}{\ed r}\right|_{\theta=\frac{\pi}{2}} =  \frac{L\left(r^2-2M r\right)-2 M a \omega_0 r}{\Delta\sqrt{\left[\omega_0 (r^2+a^2)-a L\right]^2-\Delta\left[f_\theta+f_r + (L-a \omega_0)^2\right]}}=\mathfrak{F}(r),
\end{equation}
according to which, the lens equation is written as
\begin{equation}\label{eq:lensInt}
    \hat\vartheta = 2 \int_{r_D}^{\infty} \frac{\ed r}{\mathfrak{F}(r)}-\pi,
\end{equation}
where $\hat{\vartheta}$ is the deflection angle and $r_D$ is that in Eq.~\eqref{eq:rD}. This provides 
\begin{equation}\label{eq:lesnEq}
    \hat\vartheta = 2 \tsup{\mathcal{K}}_+ \mathcal{F}_+\left(\frac{\alpha_1}{12}\right)-2 \tsup{\mathcal{K}}_- \mathcal{F}_-\left(\frac{\alpha_1}{12}\right) - 2\tsup{{B}}~ \ss\left(\frac{\alpha_1}{12}\right)-\pi,
\end{equation}
in which, $\mathcal{F}_\pm$ are given in Eq.~\eqref{eq:Fpm}, 
\begin{subequations}\label{eq:Kttpm}
\begin{align}
    & \tsup{\mathcal{K}}_\pm = \frac{r_D \left[L r_D \left(-2 a M \omega_0 (r_D-r_\pm)^2-r_D+2 r_\pm\right)-2 a M r_\pm \omega_0+L^2 r_D (r_D-2M) (r_D-r_\pm)^2-2 L M r_\pm\right]}{4 \omega_0\sqrt{\tilde{\alpha}}(r_D-r_\pm)^2(r_+-r_-)},\\
    & \tsup{B} = \frac{r_D [L (r_D-2 M)-2 a M \omega_0]}{4\omega_0 \sqrt{\tilde{\alpha}} (r_D-r_-) (r_D-r_+)},
\end{align}
\end{subequations}
and $\varUpsilon_\pm$ are the same as those in Eq.~\eqref{eq:varUpsilonpm} with $r_i\rightarrow r_D$. Applying the data given in Fig.~\ref{fig:3D}, one obtains $\hat\vartheta=47.137^\circ$ and $\hat{\vartheta} = 110.869^\circ$, respectively, for ${\omega_0 = 0.755}$ and $\omega_0=0.70$. This is while for a vacuum background (i.e. $f_r=f_\theta = 0$), these values change to $\hat\vartheta=27.276^\circ$ and $\hat\vartheta=52.997^\circ$, for the same initial frequencies.


\subsection{The evolution of the coordinate time (the $t$-motion) }

We exploit the same methods of integration, as we had for the case of the $\phi$-motion. Accordingly, considering Eq. \eqref{basiceqste}, together with Eqs.~\eqref{basiceqsR} and \eqref{basiceqstheta}, we can write the integral equation for the $t$-motion as
\begin{equation}\label{eq:h.5}
    t(\gamma) =  t_\theta(\gamma)+ t_r(\gamma),
\end{equation}
with 
\begin{subequations}\label{h.5}
\begin{align}
 &  t_\theta(\gamma) =-\int^{\theta(\gamma)}_{\theta_{\min}}
\frac{\omega_0 a^2\sin^2\theta\,\ed\theta}{\sqrt{\Theta(\theta)}}, \label{h.5a}\\
 &   t_r(\gamma) = \int^{r(\gamma)}_{r_{i}}
\frac{\left[\omega_0(r^2+a^2)^2-2MaL r\right]\ed r}{\Delta\sqrt{\mathcal{R}(r)}}. \label{h.5b}
\end{align}
\end{subequations}
Recalling the functions defined in the last subsection, the $\theta$-dependent integral above, gives these three cases:
\begin{itemize}
    \item {For $\tilde{\eta}>0$: }
\begin{equation}\label{eq:h.52}
t_{\theta}(\gamma) =   2a\Big{\{}\zeta\Big(\ss\big(U_\theta(\theta_{\min})\big)\Big)-\zeta\Big(\ss\big(U_\theta(\theta)\big)\Big)\\
+\left({1\over4}+{\chi_0^2\over 12 a^2}\right)\Big[\ss\big(U_\theta(\theta_{\min})\big)-\ss\big(U_\theta(\theta)\big)\Big]\Big{\}}.
\end{equation}

\item{For $\tilde{\eta}=0$:
\begin{equation}\label{eq:h.53}
t_{\theta}(\gamma) =  2a\Big{\{}\zeta\Big(\ss\big(\bar{U}_\theta(\theta_{\min})\big)\Big)-\zeta\Big(\ss\big(\bar{U}_\theta(\theta)\big)\Big)
+\left(\frac{1}{6}+\frac{\xi^2}{12a^2}\right)\Big[\ss\big(\bar{U}_\theta(\theta_{\min})\big)-\ss\big(\bar{U}_\theta(\theta)\big)\Big]\Big{\}}.
\end{equation}
}

\item{For $\tilde{\eta}<0$:
\begin{equation}\label{eq:h.54}
t_{\theta}(\gamma) =  2a\Big{\{}\zeta\Big(\ss\big(\bar{\bar{U}}_\theta(\theta_{\min})\big)\Big)-\zeta\Big(\ss\big(\bar{\bar{U}}_\theta(\theta)\big)\Big)
+\left(\frac{1}{4}-\frac{\mu_0^2}{12}\right)\Big[\ss\big(\bar{\bar{U}}_\theta(\theta_{\min})\big)-\ss\big(\bar{\bar{U}}_\theta(\theta)\big)\Big]\Big{\}}.
\end{equation}
}
\end{itemize}
The $r$-dependent integral \eqref{h.5b} provides the solution
\begin{equation}\label{eq:tr}
    t_r(\gamma) =\tau_0\sum_{j=1}^5 \tau_j \Big[\mathcal{T}_j\big(r(\gamma)\big)-\mathcal{T}_j(r_i)\Big],
\end{equation}
in which (letting $\mathcal{T}_j\equiv\mathcal{T}_j\big(r(\gamma)\big)$)
\begin{subequations}\label{eq:tau}
\begin{align}
&  \mathcal{T}_1 = \ss(U_r),\\
&  \mathcal{T}_2 = \frac{\wp''\left(\frac{\tilde\beta}{12}\right)}{\wp'^3\left(\frac{\tilde\beta}{12}\right)}\ln\left(
\frac{\sigma\left(\ss(U_r)+\frac{\tilde\beta}{12}\right)}{\sigma\left(\ss(U_r)-\frac{\tilde\beta}{12}\right)}\right)
-\frac{1}{\wp'^2\left(\frac{\tilde{\beta}}{12}\right)}\left[
\zeta\left(\ss(U_r)+\frac{\tilde\beta}{12}\right)
+\zeta\left(\ss(U_r)-\frac{\tilde\beta}{12}\right)\right]\nonumber\\
&\quad\quad-2\ss(U_r)\left[\frac{\wp\left(\frac{\tilde\beta}{12}\right)}{\wp'^2\left(\frac{\tilde\beta}{12}\right)}+\frac{\wp''\left(\frac{\tilde\beta}{12}\right)\zeta\left(\frac{\tilde\beta}{12}\right)}{\wp'^3\left(\frac{\tilde\beta}{12}\right)}\right],\\
&  \mathcal{T}_3  =  \frac{1}{\wp'\left(\frac{\tilde\beta}{12}\right)}\left[\ln\left(
\frac{\sigma\left(\frac{\tilde\beta}{12}-\ss(U_r)\right)}{\sigma\left(\frac{\tilde\beta}{12}+\ss(U_r)\right)}\right)
+ 2 \ss(U_r) \zeta\left(\frac{\tilde\beta}{12}\right)
\right],\\
& \mathcal{T}_4  = \mathcal{F}_+(U_r),\\
& \mathcal{T}_5  =\mathcal{F}_-(U_r),
\end{align}
\end{subequations}
recalling the expressions from the last subsection, and 
\begin{subequations}\label{eq:tau}
\begin{align}
&  \tau_0 = \frac{r_i}{\omega_0\sqrt{C_3} (r_i-r_+)(r_i-r_-)},\\
&  \tau_1 = \omega_0(r_i^2+a^2)^2-2 M a L r_i,\\
&  \tau_2 =  \frac{\omega_0}{16}\,r_i^2(r_i-r_+)(r_i-r_-),\\
&  \tau_3  = \omega_0\,r_i^4 \left[{4u_+ u_--(u_++ u_-)\over 4u_+^2 u_-^2}\right],\\
& \tau_4  = {\omega_0\left[r_i^2\left(u_+-1\right)^2+a^2u_+^2\right]^2\over 4u_+^2(u_- -u_+)}-{MaLr_i\left(u_+-1\right)u_+\over 2(u_- -u_+)},\\
& \tau_5  =- {\omega_0\left[r_i^2\left(u_--1\right)^2+a^2u_-^2\right]^2\over 4u_-^2(u_- -u_+)}+{MaLr_i\left(u_--1\right)u_-\over 2(u_- -u_+)},
\end{align}
\end{subequations}
where $u_{\pm}=\left[(r_{\pm}/r_i)-1\right]^{-1}$, the relevant Weierstra$\ss$ invariant are given in Eqs.~\eqref{eq:tg2tg3}, and the constant $C_3$ has been taken from Eq.~\eqref{eq;C3}, considering $r_D\rightarrow r_i$.

Note that, to simulate the angular trajectories, the Mino time $\gamma$ is used as the curve parameter. The above derivations for the time coordinate $t$ are, therefore, of pure mathematical significance and are presented only to have in hand the complete evolution of the spacetime coordinates. Accordingly, we pause the discussion at this point, and summarize the results in the next section.

\section{Conclusions}\label{sec:concl}

There is no doubt that physical systems offer different criteria for them to be describable efficiently and completely. In this work, we have considered such criteria for monochromatic light rays of peculiar frequency $\omega_0$, that travel in the exterior spacetime geometry of a Kerr black hole which is filled by an inhomogeneous anisotropic electronic plasma. Following the mathematical formulations founded by Synge, it is well-known that light rays will no longer travel on null geodesics inside dispersive media. We therefore, applied a proper Hamilton-Jacobi formalism which enabled us generating the differential equations of motion. Accordingly, the effective gravitational potential provides different conditions corresponding to different types of orbits for the light rays that travel on time-like trajectories with respect to the background spacetime manifold. In fact, we have tried to carry out an ambitious study on the optical conditions that govern the light propagation in plasmic medium, mostly because such studies are usually restricted to the visible limit of the black holes' exterior and their shadow. We, however, paid attention to the other types of orbits offered by the effective potential and obtained the exact analytical solutions to their respective equations of motion. We characterized the plasmic medium with the two structural functions $f_r$ and $f_{\theta}$, that enabled us conceiving the property of anisotropy. The resultant equations of motion were in the form of elliptic integrals, to solve which, we adopted specific algebraic methods that gave rise to Weierstra$\ss$ian functions as the analytical solutions. Furthermore, the dimension-less Mino time was chosen as the curve parameter, so that we could have more well-expressed coordinate evolution. This parameter was then exploited in the  parametric plots to perform three-dimensional simulations of the light ray trajectories. Note that, the effective potential does not offer any planetary orbits, so that the light rays, beside being able to form photon rings on the UCO, can only escape from or be captured by the black hole. These kinds of trajectories, as demonstrated within the text, are bounded to  cones of definite vertex angles. Based on the specific frequency (energy) of the light rays, $\omega_0$, the intensity of the deflection in the escaping trajectories can vary, and the rays may travel on more fast-changing hyperbolic curves as $\omega_0$ approaches its critical value, $\omega_U$. Note that, since we express the plasma frequency $\omega_p$ in terms of the black hole's physical characteristics in Eq.~\eqref{plasma4}, the impacts of plasma are then included, indirectly, through the variations of the effective potential and the consequent types of orbit. In the presented analytical solutions, this was done by introducing the effective impact parameter $\tilde{\eta}$ and its dependent constants. Although the temporal evolution of the coordinates have appeared to be of rather complicated mathematical expressions, they however, can establish the basement of further studies, specified to certain models of plasma that are in connection with black holes' parameters. Such studies can offer astrophysical applications and are left to our future works.

\acknowledgments

The authors would like to thank the referee for illuminating comments and precise criticism of the manuscript. M. Fathi has been supported by the Agencia Nacional de Investigaci\'{o}n y Desarrollo (ANID) through DOCTORADO Grant No. 2019-21190382, and No. 2021-242210002. J.R. Villanueva was partially supported by the Centro de Astrof\'isica de Valpara\'iso (CAV).

\appendix %


\section{Calculation of the radial deflection}\label{app:D}

Applying the method of {\it{synthetic division}} to factorize $(r-r_D)$ from the characteristic polynomial $\mathcal{P}(r)$ in Eq.~\eqref{eq:P(r)_char}, we obtain
\begin{equation}\label{eq:D3}
    \mP(r)=(r-r_D)(r^3+G_1 r^2+G_2 r+G_3),
\end{equation}
in which
\begin{subequations}\label{eq:D4}
\begin{align}
 & G_1 = r_D,\\
 & G_2 = \mA+r_D^2,\\
 & G_3 = r_D^3 + \mA r_D + \mB.
\end{align}
\end{subequations}
The deflecting trajectory can be then determined by performing, successively, the changes of variables
\begin{equation}\label{eq:D5}
    u(r) = \frac{1}{\left.\frac{r}{r_D}\right.-1},
\end{equation}
and 
\begin{equation}\label{eq:D10}
    U(r) = \frac{u(r)}{4}-\frac{6r_D^2+\mathcal{A}}{12 (4r_D^3+2\mathcal{A}r_D+\mathcal{B})}.
\end{equation}

\section{A brief review on Weierstra$\ss$ian normal elliptic integrals}\label{app:A}

Any elliptic integral of the form
\begin{equation}\label{eq:A1}
    I_\mathrm{e} = \int\frac{\mathfrak{P}(x)\,\ed x}{\sqrt{4x^3-g_2 x - g_3}},
\end{equation}
in which $\mathfrak{P}(x)$ is a rational function of $x$, can  be always evaluated, depending on
the three integrals \cite{handbookElliptic}
\begin{subequations}\label{eq:A2}
\begin{align}
    &  \int_{\wp(u)}^\infty\frac{\ed x}{\sqrt{4x^3-g_2 x - g_3}} = u, \label{eq:A2a}\\
    &  \int_{\wp(u)}^{\wp(u_0)}\frac{x \ed x}{\sqrt{4x^3-g_2 x - g_3}} = \zeta(u_0) - \zeta(u), \label{eq:A2b}\\
    &  \int_{\wp(u)}^\infty \frac{\ed x}{(x-\upsilon)\sqrt{4x^3-g_2 x - g_3}} = \int^{u}_0\frac{\ed u}{\wp(u)-\wp(\upsilon)}\nonumber\\
    &~~ =\frac{1}{\wp'(u)}\left[\ln\left(\frac{\sigma(\upsilon-u)}{\sigma(\upsilon+u)}\right) + 2 u \zeta(\upsilon)\right],\label{eq:A2c}
\end{align}
\end{subequations}
where $\wp(u)\equiv\wp(u;g_2.g_3)$, $\zeta(u)\equiv\zeta(u;g_2,g_3)$ and $\sigma(u)=\sigma(u;g_2,g_3)$, are respectively, the Weierstra$\ss$ian $\wp$, Zeta and Sigma functions, with the invariants $g_2$ and $g_3$. These functions are related together by the relations
\begin{subequations}\label{eq:A3}
\begin{align}
    & \zeta(u) = - \int\wp(u) \ed u,\\
    & \sigma(u) = \exp\left[\int\zeta(u) \ed u\right].
\end{align}
\end{subequations}
Furthermore, in Eq.~\eqref{eq:A2c}, it is defined that
\begin{equation}\label{eq:A4}
    \wp'(u) \equiv \frac{\ed}{\ed u}\wp(u) = -\sqrt{4\wp^3(u) - g_2 \wp(u) - g_3}.
\end{equation}
There is also another integral relation that we have exploited in this paper:
\begin{equation}\label{eq:A5}
   \int\frac{\ed u}{\big[\wp(u)-\wp(\upsilon)\big]^2}
   = \frac{\wp''(\upsilon)}{\wp'^3(\upsilon)}\ln\left(\frac{\sigma(u+\upsilon)}{\sigma(u-\upsilon)}\right)-\frac{1}{\wp'^2(\upsilon)}\big[\zeta(u+\upsilon)+\zeta(u-\upsilon)\big]
   -\left[\frac{2\wp(\upsilon)}{\wp'^2(\upsilon)}+\frac{2\wp''(\upsilon)\zeta(\upsilon)}{\wp'^3(\upsilon)}\right]u.
\end{equation}

%


\bibliography{biblio_v1}

\begin{thebibliography}{61}%
\makeatletter
\providecommand \@ifxundefined [1]{%
 \@ifx{#1\undefined}
}%
\providecommand \@ifnum [1]{%
 \ifnum #1\expandafter \@firstoftwo
 \else \expandafter \@secondoftwo
 \fi
}%
\providecommand \@ifx [1]{%
 \ifx #1\expandafter \@firstoftwo
 \else \expandafter \@secondoftwo
 \fi
}%
\providecommand \natexlab [1]{#1}%
\providecommand \enquote  [1]{``#1''}%
\providecommand \bibnamefont  [1]{#1}%
\providecommand \bibfnamefont [1]{#1}%
\providecommand \citenamefont [1]{#1}%
\providecommand \href@noop [0]{\@secondoftwo}%
\providecommand \href [0]{\begingroup \@sanitize@url \@href}%
\providecommand \@href[1]{\@@startlink{#1}\@@href}%
\providecommand \@@href[1]{\endgroup#1\@@endlink}%
\providecommand \@sanitize@url [0]{\catcode `\\12\catcode `\$12\catcode
  `\&12\catcode `\#12\catcode `\^12\catcode `\_12\catcode `\%12\relax}%
\providecommand \@@startlink[1]{}%
\providecommand \@@endlink[0]{}%
\providecommand \url  [0]{\begingroup\@sanitize@url \@url }%
\providecommand \@url [1]{\endgroup\@href {#1}{\urlprefix }}%
\providecommand \urlprefix  [0]{URL }%
\providecommand \Eprint [0]{\href }%
\providecommand \doibase [0]{http://dx.doi.org/}%
\providecommand \selectlanguage [0]{\@gobble}%
\providecommand \bibinfo  [0]{\@secondoftwo}%
\providecommand \bibfield  [0]{\@secondoftwo}%
\providecommand \translation [1]{[#1]}%
\providecommand \BibitemOpen [0]{}%
\providecommand \bibitemStop [0]{}%
\providecommand \bibitemNoStop [0]{.\EOS\space}%
\providecommand \EOS [0]{\spacefactor3000\relax}%
\providecommand \BibitemShut  [1]{\csname bibitem#1\endcsname}%
\let\auto@bib@innerbib\@empty
\bibitem [{\citenamefont {{Hagihara}}(1930)}]{Hagihara:1930}%
  \BibitemOpen
  \bibfield  {author} {\bibinfo {author} {\bibfnamefont {Y.}~\bibnamefont
  {{Hagihara}}},\ }\href@noop {} {\bibfield  {journal} {\bibinfo  {journal}
  {Japanese Journal of Astronomy and Geophysics}\ }\textbf {\bibinfo {volume}
  {8}},\ \bibinfo {pages} {67} (\bibinfo {year} {1930})}\BibitemShut {NoStop}%
\bibitem [{\citenamefont {{Bardeen}}\ \emph {et~al.}(1972)\citenamefont
  {{Bardeen}}, \citenamefont {{Press}},\ and\ \citenamefont
  {{Teukolsky}}}]{Bardeen:1972a}%
  \BibitemOpen
  \bibfield  {author} {\bibinfo {author} {\bibfnamefont {J.~M.}\ \bibnamefont
  {{Bardeen}}}, \bibinfo {author} {\bibfnamefont {W.~H.}\ \bibnamefont
  {{Press}}}, \ and\ \bibinfo {author} {\bibfnamefont {S.~A.}\ \bibnamefont
  {{Teukolsky}}},\ }\href {\doibase 10.1086/151796} {\bibfield  {journal}
  {\bibinfo  {journal} {Astrophysical Journal}\ }\textbf {\bibinfo {volume}
  {178}},\ \bibinfo {pages} {347} (\bibinfo {year} {1972})}\BibitemShut
  {NoStop}%
\bibitem [{\citenamefont {Bardeen}(1973)}]{Bardeen:1973a}%
  \BibitemOpen
  \bibfield  {author} {\bibinfo {author} {\bibfnamefont {J.}~\bibnamefont
  {Bardeen}},\ }in\ \href@noop {} {\emph {\bibinfo {booktitle} {{Les Houches
  Summer School of Theoretical Physics}: {Black Holes}}}}\ (\bibinfo {year}
  {1973})\ pp.\ \bibinfo {pages} {215--240}\BibitemShut {NoStop}%
\bibitem [{\citenamefont {{Cunningham}}\ and\ \citenamefont
  {{Bardeen}}(1973)}]{Bardeen:1973b}%
  \BibitemOpen
  \bibfield  {author} {\bibinfo {author} {\bibfnamefont {C.~T.}\ \bibnamefont
  {{Cunningham}}}\ and\ \bibinfo {author} {\bibfnamefont {J.~M.}\ \bibnamefont
  {{Bardeen}}},\ }\href {\doibase 10.1086/152223} {\bibfield  {journal}
  {\bibinfo  {journal} {Astrophysical Journal}\ }\textbf {\bibinfo {volume}
  {183}},\ \bibinfo {pages} {237} (\bibinfo {year} {1973})}\BibitemShut
  {NoStop}%
\bibitem [{\citenamefont {Chandrasekhar}(2002)}]{Chandrasekhar:2002}%
  \BibitemOpen
  \bibfield  {author} {\bibinfo {author} {\bibfnamefont {S.}~\bibnamefont
  {Chandrasekhar}},\ }\href {https://cds.cern.ch/record/579245} {\emph
  {\bibinfo {title} {The mathematical theory of black holes}}},\ Oxford classic
  texts in the physical sciences\ (\bibinfo  {publisher} {Oxford Univ. Press},\
  \bibinfo {address} {Oxford},\ \bibinfo {year} {2002})\BibitemShut {NoStop}%
\bibitem [{\citenamefont {Virbhadra}\ and\ \citenamefont
  {Ellis}(2000)}]{Virbhadra:2000}%
  \BibitemOpen
  \bibfield  {author} {\bibinfo {author} {\bibfnamefont {K.~S.}\ \bibnamefont
  {Virbhadra}}\ and\ \bibinfo {author} {\bibfnamefont {G.~F.~R.}\ \bibnamefont
  {Ellis}},\ }\href {\doibase 10.1103/PhysRevD.62.084003} {\bibfield  {journal}
  {\bibinfo  {journal} {Physical Review D}\ }\textbf {\bibinfo {volume} {62}},\
  \bibinfo {pages} {084003} (\bibinfo {year} {2000})}\BibitemShut {NoStop}%
\bibitem [{\citenamefont {Kraniotis}(2004)}]{kraniotis_precise_2004}%
  \BibitemOpen
  \bibfield  {author} {\bibinfo {author} {\bibfnamefont {G.~V.}\ \bibnamefont
  {Kraniotis}},\ }\href {\doibase 10.1088/0264-9381/21/19/016} {\bibfield
  {journal} {\bibinfo  {journal} {Classical and Quantum Gravity}\ }\textbf
  {\bibinfo {volume} {21}},\ \bibinfo {pages} {4743} (\bibinfo {year}
  {2004})}\BibitemShut {NoStop}%
\bibitem [{\citenamefont {Beckwith}\ and\ \citenamefont
  {Done}(2005)}]{beckwith_extreme_2005}%
  \BibitemOpen
  \bibfield  {author} {\bibinfo {author} {\bibfnamefont {K.}~\bibnamefont
  {Beckwith}}\ and\ \bibinfo {author} {\bibfnamefont {C.}~\bibnamefont
  {Done}},\ }\href {\doibase 10.1111/j.1365-2966.2005.08980.x} {\bibfield
  {journal} {\bibinfo  {journal} {Monthly Notices of the Royal Astronomical
  Society}\ }\textbf {\bibinfo {volume} {359}},\ \bibinfo {pages} {1217}
  (\bibinfo {year} {2005})}\BibitemShut {NoStop}%
\bibitem [{\citenamefont {Kraniotis}(2005)}]{kraniotis_frame_2005}%
  \BibitemOpen
  \bibfield  {author} {\bibinfo {author} {\bibfnamefont {G.~V.}\ \bibnamefont
  {Kraniotis}},\ }\href {\doibase 10.1088/0264-9381/22/21/001} {\bibfield
  {journal} {\bibinfo  {journal} {Classical and Quantum Gravity}\ }\textbf
  {\bibinfo {volume} {22}},\ \bibinfo {pages} {4391} (\bibinfo {year}
  {2005})}\BibitemShut {NoStop}%
\bibitem [{\citenamefont {Hackmann}\ and\ \citenamefont
  {Lämmerzahl}(2008{\natexlab{a}})}]{hackmann_complete_2008}%
  \BibitemOpen
  \bibfield  {author} {\bibinfo {author} {\bibfnamefont {E.}~\bibnamefont
  {Hackmann}}\ and\ \bibinfo {author} {\bibfnamefont {C.}~\bibnamefont
  {Lämmerzahl}},\ }\href {\doibase 10.1103/PhysRevLett.100.171101} {\bibfield
  {journal} {\bibinfo  {journal} {Physical Review Letters}\ }\textbf {\bibinfo
  {volume} {100}},\ \bibinfo {pages} {171101} (\bibinfo {year}
  {2008}{\natexlab{a}})}\BibitemShut {NoStop}%
\bibitem [{\citenamefont {Hackmann}\ and\ \citenamefont
  {Lämmerzahl}(2008{\natexlab{b}})}]{hackmann_geodesic_2008}%
  \BibitemOpen
  \bibfield  {author} {\bibinfo {author} {\bibfnamefont {E.}~\bibnamefont
  {Hackmann}}\ and\ \bibinfo {author} {\bibfnamefont {C.}~\bibnamefont
  {Lämmerzahl}},\ }\href {\doibase 10.1103/PhysRevD.78.024035} {\bibfield
  {journal} {\bibinfo  {journal} {Physical Review D}\ }\textbf {\bibinfo
  {volume} {78}},\ \bibinfo {pages} {024035} (\bibinfo {year}
  {2008}{\natexlab{b}})}\BibitemShut {NoStop}%
\bibitem [{\citenamefont {Bisnovatyi-Kogan}\ and\ \citenamefont
  {Tsupko}(2008)}]{bisnovatyi-kogan_strong_2008}%
  \BibitemOpen
  \bibfield  {author} {\bibinfo {author} {\bibfnamefont {G.~S.}\ \bibnamefont
  {Bisnovatyi-Kogan}}\ and\ \bibinfo {author} {\bibfnamefont {O.~Y.}\
  \bibnamefont {Tsupko}},\ }\href {\doibase 10.1007/s10511-008-0011-8}
  {\bibfield  {journal} {\bibinfo  {journal} {Astrophysics}\ }\textbf {\bibinfo
  {volume} {51}},\ \bibinfo {pages} {99} (\bibinfo {year} {2008})}\BibitemShut
  {NoStop}%
\bibitem [{\citenamefont {Kagramanova}\ \emph {et~al.}(2010)\citenamefont
  {Kagramanova}, \citenamefont {Kunz}, \citenamefont {Hackmann},\ and\
  \citenamefont {Lämmerzahl}}]{kagramanova_analytic_2010}%
  \BibitemOpen
  \bibfield  {author} {\bibinfo {author} {\bibfnamefont {V.}~\bibnamefont
  {Kagramanova}}, \bibinfo {author} {\bibfnamefont {J.}~\bibnamefont {Kunz}},
  \bibinfo {author} {\bibfnamefont {E.}~\bibnamefont {Hackmann}}, \ and\
  \bibinfo {author} {\bibfnamefont {C.}~\bibnamefont {Lämmerzahl}},\ }\href
  {\doibase 10.1103/PhysRevD.81.124044} {\bibfield  {journal} {\bibinfo
  {journal} {Physical Review D}\ }\textbf {\bibinfo {volume} {81}},\ \bibinfo
  {pages} {124044} (\bibinfo {year} {2010})}\BibitemShut {NoStop}%
\bibitem [{\citenamefont {Hackmann}\ \emph
  {et~al.}(2010{\natexlab{a}})\citenamefont {Hackmann}, \citenamefont
  {Lämmerzahl}, \citenamefont {Kagramanova},\ and\ \citenamefont
  {Kunz}}]{hackmann_analytical_2010}%
  \BibitemOpen
  \bibfield  {author} {\bibinfo {author} {\bibfnamefont {E.}~\bibnamefont
  {Hackmann}}, \bibinfo {author} {\bibfnamefont {C.}~\bibnamefont
  {Lämmerzahl}}, \bibinfo {author} {\bibfnamefont {V.}~\bibnamefont
  {Kagramanova}}, \ and\ \bibinfo {author} {\bibfnamefont {J.}~\bibnamefont
  {Kunz}},\ }\href {\doibase 10.1103/PhysRevD.81.044020} {\bibfield  {journal}
  {\bibinfo  {journal} {Physical Review D}\ }\textbf {\bibinfo {volume} {81}},\
  \bibinfo {pages} {044020} (\bibinfo {year} {2010}{\natexlab{a}})}\BibitemShut
  {NoStop}%
\bibitem [{\citenamefont {Hackmann}\ \emph
  {et~al.}(2010{\natexlab{b}})\citenamefont {Hackmann}, \citenamefont
  {Hartmann}, \citenamefont {Lämmerzahl},\ and\ \citenamefont
  {Sirimachan}}]{hackmann_complete_2010}%
  \BibitemOpen
  \bibfield  {author} {\bibinfo {author} {\bibfnamefont {E.}~\bibnamefont
  {Hackmann}}, \bibinfo {author} {\bibfnamefont {B.}~\bibnamefont {Hartmann}},
  \bibinfo {author} {\bibfnamefont {C.}~\bibnamefont {Lämmerzahl}}, \ and\
  \bibinfo {author} {\bibfnamefont {P.}~\bibnamefont {Sirimachan}},\ }\href
  {\doibase 10.1103/PhysRevD.81.064016} {\bibfield  {journal} {\bibinfo
  {journal} {Physical Review D}\ }\textbf {\bibinfo {volume} {81}},\ \bibinfo
  {pages} {064016} (\bibinfo {year} {2010}{\natexlab{b}})}\BibitemShut
  {NoStop}%
\bibitem [{\citenamefont {Enolski}\ \emph {et~al.}(2011)\citenamefont
  {Enolski}, \citenamefont {Hackmann}, \citenamefont {Kagramanova},
  \citenamefont {Kunz},\ and\ \citenamefont
  {Lämmerzahl}}]{enolski_inversion_2011}%
  \BibitemOpen
  \bibfield  {author} {\bibinfo {author} {\bibfnamefont {V.}~\bibnamefont
  {Enolski}}, \bibinfo {author} {\bibfnamefont {E.}~\bibnamefont {Hackmann}},
  \bibinfo {author} {\bibfnamefont {V.}~\bibnamefont {Kagramanova}}, \bibinfo
  {author} {\bibfnamefont {J.}~\bibnamefont {Kunz}}, \ and\ \bibinfo {author}
  {\bibfnamefont {C.}~\bibnamefont {Lämmerzahl}},\ }\href {\doibase
  10.1016/j.geomphys.2011.01.001} {\bibfield  {journal} {\bibinfo  {journal}
  {Journal of Geometry and Physics}\ }\textbf {\bibinfo {volume} {61}},\
  \bibinfo {pages} {899} (\bibinfo {year} {2011})}\BibitemShut {NoStop}%
\bibitem [{\citenamefont {Kraniotis}(2011)}]{kraniotis_precise_2011}%
  \BibitemOpen
  \bibfield  {author} {\bibinfo {author} {\bibfnamefont {G.~V.}\ \bibnamefont
  {Kraniotis}},\ }\href {\doibase 10.1088/0264-9381/28/8/085021} {\bibfield
  {journal} {\bibinfo  {journal} {Classical and Quantum Gravity}\ }\textbf
  {\bibinfo {volume} {28}},\ \bibinfo {pages} {085021} (\bibinfo {year}
  {2011})}\BibitemShut {NoStop}%
\bibitem [{\citenamefont {Enolski}\ \emph {et~al.}(2012)\citenamefont
  {Enolski}, \citenamefont {Hartmann}, \citenamefont {Kagramanova},
  \citenamefont {Kunz}, \citenamefont {Lämmerzahl},\ and\ \citenamefont
  {Sirimachan}}]{enolski_inversion_2012}%
  \BibitemOpen
  \bibfield  {author} {\bibinfo {author} {\bibfnamefont {V.}~\bibnamefont
  {Enolski}}, \bibinfo {author} {\bibfnamefont {B.}~\bibnamefont {Hartmann}},
  \bibinfo {author} {\bibfnamefont {V.}~\bibnamefont {Kagramanova}}, \bibinfo
  {author} {\bibfnamefont {J.}~\bibnamefont {Kunz}}, \bibinfo {author}
  {\bibfnamefont {C.}~\bibnamefont {Lämmerzahl}}, \ and\ \bibinfo {author}
  {\bibfnamefont {P.}~\bibnamefont {Sirimachan}},\ }\href {\doibase
  10.1063/1.3677831} {\bibfield  {journal} {\bibinfo  {journal} {Journal of
  Mathematical Physics}\ }\textbf {\bibinfo {volume} {53}},\ \bibinfo {pages}
  {012504} (\bibinfo {year} {2012})}\BibitemShut {NoStop}%
\bibitem [{\citenamefont {Gibbons}\ and\ \citenamefont
  {Vyska}(2012)}]{gibbons_application_2012}%
  \BibitemOpen
  \bibfield  {author} {\bibinfo {author} {\bibfnamefont {G.~W.}\ \bibnamefont
  {Gibbons}}\ and\ \bibinfo {author} {\bibfnamefont {M.}~\bibnamefont
  {Vyska}},\ }\href {\doibase 10.1088/0264-9381/29/6/065016} {\bibfield
  {journal} {\bibinfo  {journal} {Classical and Quantum Gravity}\ }\textbf
  {\bibinfo {volume} {29}},\ \bibinfo {pages} {065016} (\bibinfo {year}
  {2012})}\BibitemShut {NoStop}%
\bibitem [{\citenamefont {Muñoz}(2014)}]{munoz_orbits_2014}%
  \BibitemOpen
  \bibfield  {author} {\bibinfo {author} {\bibfnamefont {G.}~\bibnamefont
  {Muñoz}},\ }\href {\doibase 10.1119/1.4866274} {\bibfield  {journal}
  {\bibinfo  {journal} {American Journal of Physics}\ }\textbf {\bibinfo
  {volume} {82}},\ \bibinfo {pages} {564} (\bibinfo {year} {2014})}\BibitemShut
  {NoStop}%
\bibitem [{\citenamefont {Kraniotis}(2014)}]{kraniotis_gravitational_2014}%
  \BibitemOpen
  \bibfield  {author} {\bibinfo {author} {\bibfnamefont {G.~V.}\ \bibnamefont
  {Kraniotis}},\ }\href {\doibase 10.1007/s10714-014-1818-8} {\bibfield
  {journal} {\bibinfo  {journal} {General Relativity and Gravitation}\ }\textbf
  {\bibinfo {volume} {46}},\ \bibinfo {pages} {1818} (\bibinfo {year}
  {2014})}\BibitemShut {NoStop}%
\bibitem [{\citenamefont {De~Falco}\ \emph {et~al.}(2016)\citenamefont
  {De~Falco}, \citenamefont {Falanga},\ and\ \citenamefont
  {Stella}}]{de_falco_approximate_2016}%
  \BibitemOpen
  \bibfield  {author} {\bibinfo {author} {\bibfnamefont {V.}~\bibnamefont
  {De~Falco}}, \bibinfo {author} {\bibfnamefont {M.}~\bibnamefont {Falanga}}, \
  and\ \bibinfo {author} {\bibfnamefont {L.}~\bibnamefont {Stella}},\ }\href
  {\doibase 10.1051/0004-6361/201629075} {\bibfield  {journal} {\bibinfo
  {journal} {Astronomy \& Astrophysics}\ }\textbf {\bibinfo {volume} {595}},\
  \bibinfo {pages} {A38} (\bibinfo {year} {2016})}\BibitemShut {NoStop}%
\bibitem [{\citenamefont {Soroushfar}\ \emph {et~al.}(2016)\citenamefont
  {Soroushfar}, \citenamefont {Saffari}, \citenamefont {Kazempour},
  \citenamefont {Grunau},\ and\ \citenamefont
  {Kunz}}]{soroushfar_detailed_2016}%
  \BibitemOpen
  \bibfield  {author} {\bibinfo {author} {\bibfnamefont {S.}~\bibnamefont
  {Soroushfar}}, \bibinfo {author} {\bibfnamefont {R.}~\bibnamefont {Saffari}},
  \bibinfo {author} {\bibfnamefont {S.}~\bibnamefont {Kazempour}}, \bibinfo
  {author} {\bibfnamefont {S.}~\bibnamefont {Grunau}}, \ and\ \bibinfo {author}
  {\bibfnamefont {J.}~\bibnamefont {Kunz}},\ }\href {\doibase
  10.1103/PhysRevD.94.024052} {\bibfield  {journal} {\bibinfo  {journal}
  {Physical Review D}\ }\textbf {\bibinfo {volume} {94}},\ \bibinfo {pages}
  {024052} (\bibinfo {year} {2016})}\BibitemShut {NoStop}%
\bibitem [{\citenamefont {Barlow}\ \emph {et~al.}(2017)\citenamefont {Barlow},
  \citenamefont {Weinstein},\ and\ \citenamefont
  {Faber}}]{barlow_asymptotically_2017}%
  \BibitemOpen
  \bibfield  {author} {\bibinfo {author} {\bibfnamefont {N.~S.}\ \bibnamefont
  {Barlow}}, \bibinfo {author} {\bibfnamefont {S.~J.}\ \bibnamefont
  {Weinstein}}, \ and\ \bibinfo {author} {\bibfnamefont {J.~A.}\ \bibnamefont
  {Faber}},\ }\href {\doibase 10.1088/1361-6382/aa7538} {\bibfield  {journal}
  {\bibinfo  {journal} {Classical and Quantum Gravity}\ }\textbf {\bibinfo
  {volume} {34}},\ \bibinfo {pages} {135017} (\bibinfo {year}
  {2017})}\BibitemShut {NoStop}%
\bibitem [{\citenamefont {Uniyal}\ \emph {et~al.}(2018)\citenamefont {Uniyal},
  \citenamefont {Nandan},\ and\ \citenamefont {Purohit}}]{uniyal_null_2018}%
  \BibitemOpen
  \bibfield  {author} {\bibinfo {author} {\bibfnamefont {R.}~\bibnamefont
  {Uniyal}}, \bibinfo {author} {\bibfnamefont {H.}~\bibnamefont {Nandan}}, \
  and\ \bibinfo {author} {\bibfnamefont {K.~D.}\ \bibnamefont {Purohit}},\
  }\href {\doibase 10.1088/1361-6382/aa9ad9} {\bibfield  {journal} {\bibinfo
  {journal} {Classical and Quantum Gravity}\ }\textbf {\bibinfo {volume}
  {35}},\ \bibinfo {pages} {025003} (\bibinfo {year} {2018})}\BibitemShut
  {NoStop}%
\bibitem [{\citenamefont {Villanueva}\ \emph {et~al.}(2018)\citenamefont
  {Villanueva}, \citenamefont {Tapia}, \citenamefont {Molina},\ and\
  \citenamefont {Olivares}}]{villanueva_null_2018}%
  \BibitemOpen
  \bibfield  {author} {\bibinfo {author} {\bibfnamefont {J.~R.}\ \bibnamefont
  {Villanueva}}, \bibinfo {author} {\bibfnamefont {F.}~\bibnamefont {Tapia}},
  \bibinfo {author} {\bibfnamefont {M.}~\bibnamefont {Molina}}, \ and\ \bibinfo
  {author} {\bibfnamefont {M.}~\bibnamefont {Olivares}},\ }\href {\doibase
  10.1140/epjc/s10052-018-6328-5} {\bibfield  {journal} {\bibinfo  {journal}
  {The European Physical Journal C}\ }\textbf {\bibinfo {volume} {78}},\
  \bibinfo {pages} {853} (\bibinfo {year} {2018})}\BibitemShut {NoStop}%
\bibitem [{\citenamefont {Chatterjee}\ \emph {et~al.}(2019)\citenamefont
  {Chatterjee}, \citenamefont {Flathmann}, \citenamefont {Nandan},\ and\
  \citenamefont {Rudra}}]{chatterjee_analytic_2019}%
  \BibitemOpen
  \bibfield  {author} {\bibinfo {author} {\bibfnamefont {A.~K.}\ \bibnamefont
  {Chatterjee}}, \bibinfo {author} {\bibfnamefont {K.}~\bibnamefont
  {Flathmann}}, \bibinfo {author} {\bibfnamefont {H.}~\bibnamefont {Nandan}}, \
  and\ \bibinfo {author} {\bibfnamefont {A.}~\bibnamefont {Rudra}},\ }\href
  {\doibase 10.1103/PhysRevD.100.024044} {\bibfield  {journal} {\bibinfo
  {journal} {Physical Review D}\ }\textbf {\bibinfo {volume} {100}},\ \bibinfo
  {pages} {024044} (\bibinfo {year} {2019})}\BibitemShut {NoStop}%
\bibitem [{\citenamefont {Hsiao}\ \emph {et~al.}(2020)\citenamefont {Hsiao},
  \citenamefont {Lee},\ and\ \citenamefont {Lin}}]{hsiao_equatorial_2020}%
  \BibitemOpen
  \bibfield  {author} {\bibinfo {author} {\bibfnamefont {Y.-W.}\ \bibnamefont
  {Hsiao}}, \bibinfo {author} {\bibfnamefont {D.-S.}\ \bibnamefont {Lee}}, \
  and\ \bibinfo {author} {\bibfnamefont {C.-Y.}\ \bibnamefont {Lin}},\ }\href
  {\doibase 10.1103/PhysRevD.101.064070} {\bibfield  {journal} {\bibinfo
  {journal} {Physical Review D}\ }\textbf {\bibinfo {volume} {101}},\ \bibinfo
  {pages} {064070} (\bibinfo {year} {2020})}\BibitemShut {NoStop}%
\bibitem [{\citenamefont {Gralla}\ and\ \citenamefont
  {Lupsasca}(2020)}]{gralla_null_2020}%
  \BibitemOpen
  \bibfield  {author} {\bibinfo {author} {\bibfnamefont {S.~E.}\ \bibnamefont
  {Gralla}}\ and\ \bibinfo {author} {\bibfnamefont {A.}~\bibnamefont
  {Lupsasca}},\ }\href {\doibase 10.1103/PhysRevD.101.044032} {\bibfield
  {journal} {\bibinfo  {journal} {Physical Review D}\ }\textbf {\bibinfo
  {volume} {101}},\ \bibinfo {pages} {044032} (\bibinfo {year}
  {2020})}\BibitemShut {NoStop}%
\bibitem [{\citenamefont {Hendi}\ \emph {et~al.}(2020)\citenamefont {Hendi},
  \citenamefont {Tavakkoli}, \citenamefont {Panahiyan}, \citenamefont {Panah},\
  and\ \citenamefont {Hackmann}}]{hendi_simulation_2020}%
  \BibitemOpen
  \bibfield  {author} {\bibinfo {author} {\bibfnamefont {S.~H.}\ \bibnamefont
  {Hendi}}, \bibinfo {author} {\bibfnamefont {A.~M.}\ \bibnamefont
  {Tavakkoli}}, \bibinfo {author} {\bibfnamefont {S.}~\bibnamefont
  {Panahiyan}}, \bibinfo {author} {\bibfnamefont {B.~E.}\ \bibnamefont
  {Panah}}, \ and\ \bibinfo {author} {\bibfnamefont {E.}~\bibnamefont
  {Hackmann}},\ }\href {\doibase 10.1140/epjc/s10052-020-8065-9} {\bibfield
  {journal} {\bibinfo  {journal} {The European Physical Journal C}\ }\textbf
  {\bibinfo {volume} {80}},\ \bibinfo {pages} {524} (\bibinfo {year}
  {2020})}\BibitemShut {NoStop}%
\bibitem [{\citenamefont {Fathi}\ \emph {et~al.}(2020)\citenamefont {Fathi},
  \citenamefont {Olivares},\ and\ \citenamefont
  {Villanueva}}]{fathi_classical_2020}%
  \BibitemOpen
  \bibfield  {author} {\bibinfo {author} {\bibfnamefont {M.}~\bibnamefont
  {Fathi}}, \bibinfo {author} {\bibfnamefont {M.}~\bibnamefont {Olivares}}, \
  and\ \bibinfo {author} {\bibfnamefont {J.~R.}\ \bibnamefont {Villanueva}},\
  }\href {\doibase 10.1140/epjc/s10052-020-7623-5} {\bibfield  {journal}
  {\bibinfo  {journal} {The European Physical Journal C}\ }\textbf {\bibinfo
  {volume} {80}},\ \bibinfo {pages} {51} (\bibinfo {year} {2020})}\BibitemShut
  {NoStop}%
\bibitem [{\citenamefont {Kraniotis}(2021)}]{kraniotis_gravitational_2021}%
  \BibitemOpen
  \bibfield  {author} {\bibinfo {author} {\bibfnamefont {G.~V.}\ \bibnamefont
  {Kraniotis}},\ }\href {\doibase 10.1140/epjc/s10052-021-08911-5} {\bibfield
  {journal} {\bibinfo  {journal} {The European Physical Journal C}\ }\textbf
  {\bibinfo {volume} {81}},\ \bibinfo {pages} {147} (\bibinfo {year}
  {2021})}\BibitemShut {NoStop}%
\bibitem [{\citenamefont {{Synge}}(1960)}]{Synge:1960}%
  \BibitemOpen
  \bibfield  {author} {\bibinfo {author} {\bibfnamefont {J.~L.}\ \bibnamefont
  {{Synge}}},\ }\href@noop {} {\emph {\bibinfo {title} {{Relativity: The
  general theory}}}}\ (\bibinfo  {publisher} {Series in Physics, Amsterdam:
  North-Holland Publication Co.},\ \bibinfo {year} {1960})\BibitemShut
  {NoStop}%
\bibitem [{\citenamefont {Bisnovatyi-Kogan}\ and\ \citenamefont
  {Tsupko}(2009)}]{bisnovatyi-kogan_gravitational_2009}%
  \BibitemOpen
  \bibfield  {author} {\bibinfo {author} {\bibfnamefont {G.~S.}\ \bibnamefont
  {Bisnovatyi-Kogan}}\ and\ \bibinfo {author} {\bibfnamefont {O.~Y.}\
  \bibnamefont {Tsupko}},\ }\href {\doibase 10.1134/S020228930901006X}
  {\bibfield  {journal} {\bibinfo  {journal} {Gravitation and Cosmology}\
  }\textbf {\bibinfo {volume} {15}},\ \bibinfo {pages} {20} (\bibinfo {year}
  {2009})}\BibitemShut {NoStop}%
\bibitem [{\citenamefont {Bisnovatyi-Kogan}\ and\ \citenamefont
  {Tsupko}(2010)}]{bisnovatyi-kogan_gravitational_2010}%
  \BibitemOpen
  \bibfield  {author} {\bibinfo {author} {\bibfnamefont {G.~S.}\ \bibnamefont
  {Bisnovatyi-Kogan}}\ and\ \bibinfo {author} {\bibfnamefont {O.~Y.}\
  \bibnamefont {Tsupko}},\ }\href {\doibase 10.1111/j.1365-2966.2010.16290.x}
  {\bibfield  {journal} {\bibinfo  {journal} {Monthly Notices of the Royal
  Astronomical Society}\ ,\ \bibinfo {pages} {no}} (\bibinfo {year}
  {2010})}\BibitemShut {NoStop}%
\bibitem [{\citenamefont {Tsupko}\ and\ \citenamefont
  {Bisnovatyi-Kogan}(2013)}]{tsupko_gravitational_2013}%
  \BibitemOpen
  \bibfield  {author} {\bibinfo {author} {\bibfnamefont {O.~Y.}\ \bibnamefont
  {Tsupko}}\ and\ \bibinfo {author} {\bibfnamefont {G.~S.}\ \bibnamefont
  {Bisnovatyi-Kogan}},\ }\href {\doibase 10.1103/PhysRevD.87.124009} {\bibfield
   {journal} {\bibinfo  {journal} {Physical Review D}\ }\textbf {\bibinfo
  {volume} {87}},\ \bibinfo {pages} {124009} (\bibinfo {year}
  {2013})}\BibitemShut {NoStop}%
\bibitem [{\citenamefont {Morozova}\ \emph {et~al.}(2013)\citenamefont
  {Morozova}, \citenamefont {Ahmedov},\ and\ \citenamefont
  {Tursunov}}]{morozova_gravitational_2013}%
  \BibitemOpen
  \bibfield  {author} {\bibinfo {author} {\bibfnamefont {V.~S.}\ \bibnamefont
  {Morozova}}, \bibinfo {author} {\bibfnamefont {B.~J.}\ \bibnamefont
  {Ahmedov}}, \ and\ \bibinfo {author} {\bibfnamefont {A.~A.}\ \bibnamefont
  {Tursunov}},\ }\href {\doibase 10.1007/s10509-013-1458-6} {\bibfield
  {journal} {\bibinfo  {journal} {Astrophysics and Space Science}\ }\textbf
  {\bibinfo {volume} {346}},\ \bibinfo {pages} {513} (\bibinfo {year}
  {2013})}\BibitemShut {NoStop}%
\bibitem [{\citenamefont {Bisnovatyi-Kogan}\ and\ \citenamefont
  {Tsupko}(2015)}]{bisnovatyi-kogan_gravitational_2015}%
  \BibitemOpen
  \bibfield  {author} {\bibinfo {author} {\bibfnamefont {G.~S.}\ \bibnamefont
  {Bisnovatyi-Kogan}}\ and\ \bibinfo {author} {\bibfnamefont {O.~Y.}\
  \bibnamefont {Tsupko}},\ }\href {\doibase 10.1134/S1063780X15070016}
  {\bibfield  {journal} {\bibinfo  {journal} {Plasma Physics Reports}\ }\textbf
  {\bibinfo {volume} {41}},\ \bibinfo {pages} {562} (\bibinfo {year}
  {2015})}\BibitemShut {NoStop}%
\bibitem [{\citenamefont {Perlick}\ \emph {et~al.}(2015)\citenamefont
  {Perlick}, \citenamefont {Tsupko},\ and\ \citenamefont
  {Bisnovatyi-Kogan}}]{perlick_influence_2015}%
  \BibitemOpen
  \bibfield  {author} {\bibinfo {author} {\bibfnamefont {V.}~\bibnamefont
  {Perlick}}, \bibinfo {author} {\bibfnamefont {O.~Y.}\ \bibnamefont {Tsupko}},
  \ and\ \bibinfo {author} {\bibfnamefont {G.~S.}\ \bibnamefont
  {Bisnovatyi-Kogan}},\ }\href {\doibase 10.1103/PhysRevD.92.104031} {\bibfield
   {journal} {\bibinfo  {journal} {Physical Review D}\ }\textbf {\bibinfo
  {volume} {92}},\ \bibinfo {pages} {104031} (\bibinfo {year}
  {2015})}\BibitemShut {NoStop}%
\bibitem [{\citenamefont {Atamurotov}\ \emph
  {et~al.}(2015{\natexlab{a}})\citenamefont {Atamurotov}, \citenamefont
  {Ahmedov},\ and\ \citenamefont {Abdujabbarov}}]{atamurotov_optical_2015}%
  \BibitemOpen
  \bibfield  {author} {\bibinfo {author} {\bibfnamefont {F.}~\bibnamefont
  {Atamurotov}}, \bibinfo {author} {\bibfnamefont {B.}~\bibnamefont {Ahmedov}},
  \ and\ \bibinfo {author} {\bibfnamefont {A.}~\bibnamefont {Abdujabbarov}},\
  }\href {\doibase 10.1103/PhysRevD.92.084005} {\bibfield  {journal} {\bibinfo
  {journal} {Physical Review D}\ }\textbf {\bibinfo {volume} {92}},\ \bibinfo
  {pages} {084005} (\bibinfo {year} {2015}{\natexlab{a}})}\BibitemShut
  {NoStop}%
\bibitem [{\citenamefont {Abdujabbarov}\ \emph {et~al.}(2016)\citenamefont
  {Abdujabbarov}, \citenamefont {Amir}, \citenamefont {Ahmedov},\ and\
  \citenamefont {Ghosh}}]{abdujabbarov_shadow_2016}%
  \BibitemOpen
  \bibfield  {author} {\bibinfo {author} {\bibfnamefont {A.}~\bibnamefont
  {Abdujabbarov}}, \bibinfo {author} {\bibfnamefont {M.}~\bibnamefont {Amir}},
  \bibinfo {author} {\bibfnamefont {B.}~\bibnamefont {Ahmedov}}, \ and\
  \bibinfo {author} {\bibfnamefont {S.~G.}\ \bibnamefont {Ghosh}},\ }\href
  {\doibase 10.1103/PhysRevD.93.104004} {\bibfield  {journal} {\bibinfo
  {journal} {Physical Review D}\ }\textbf {\bibinfo {volume} {93}},\ \bibinfo
  {pages} {104004} (\bibinfo {year} {2016})}\BibitemShut {NoStop}%
\bibitem [{\citenamefont {Bisnovatyi-Kogan}\ and\ \citenamefont
  {Tsupko}(2017{\natexlab{a}})}]{bisnovatyi-kogan_gravitational_2017}%
  \BibitemOpen
  \bibfield  {author} {\bibinfo {author} {\bibfnamefont {G.}~\bibnamefont
  {Bisnovatyi-Kogan}}\ and\ \bibinfo {author} {\bibfnamefont {O.}~\bibnamefont
  {Tsupko}},\ }\href {\doibase 10.3390/universe3030057} {\bibfield  {journal}
  {\bibinfo  {journal} {Universe}\ }\textbf {\bibinfo {volume} {3}},\ \bibinfo
  {pages} {57} (\bibinfo {year} {2017}{\natexlab{a}})}\BibitemShut {NoStop}%
\bibitem [{\citenamefont {Perlick}\ and\ \citenamefont
  {Tsupko}(2017{\natexlab{a}})}]{perlick_light_2017}%
  \BibitemOpen
  \bibfield  {author} {\bibinfo {author} {\bibfnamefont {V.}~\bibnamefont
  {Perlick}}\ and\ \bibinfo {author} {\bibfnamefont {O.~Y.}\ \bibnamefont
  {Tsupko}},\ }\href {\doibase 10.1103/PhysRevD.95.104003} {\bibfield
  {journal} {\bibinfo  {journal} {Physical Review D}\ }\textbf {\bibinfo
  {volume} {95}},\ \bibinfo {pages} {104003} (\bibinfo {year}
  {2017}{\natexlab{a}})}\BibitemShut {NoStop}%
\bibitem [{\citenamefont {Schulze-Koops}\ \emph {et~al.}(2017)\citenamefont
  {Schulze-Koops}, \citenamefont {Perlick},\ and\ \citenamefont
  {Schwarz}}]{schulze-koops_sachs_2017}%
  \BibitemOpen
  \bibfield  {author} {\bibinfo {author} {\bibfnamefont {K.}~\bibnamefont
  {Schulze-Koops}}, \bibinfo {author} {\bibfnamefont {V.}~\bibnamefont
  {Perlick}}, \ and\ \bibinfo {author} {\bibfnamefont {D.~J.}\ \bibnamefont
  {Schwarz}},\ }\href {\doibase 10.1088/1361-6382/aa8d46} {\bibfield  {journal}
  {\bibinfo  {journal} {Classical and Quantum Gravity}\ }\textbf {\bibinfo
  {volume} {34}},\ \bibinfo {pages} {215006} (\bibinfo {year}
  {2017})}\BibitemShut {NoStop}%
\bibitem [{\citenamefont {Abdujabbarov}\ \emph {et~al.}(2017)\citenamefont
  {Abdujabbarov}, \citenamefont {Toshmatov}, \citenamefont {Stuchlík},\ and\
  \citenamefont {Ahmedov}}]{abdujabbarov_shadow_2017}%
  \BibitemOpen
  \bibfield  {author} {\bibinfo {author} {\bibfnamefont {A.}~\bibnamefont
  {Abdujabbarov}}, \bibinfo {author} {\bibfnamefont {B.}~\bibnamefont
  {Toshmatov}}, \bibinfo {author} {\bibfnamefont {Z.}~\bibnamefont
  {Stuchlík}}, \ and\ \bibinfo {author} {\bibfnamefont {B.}~\bibnamefont
  {Ahmedov}},\ }\href {\doibase 10.1142/S0218271817500511} {\bibfield
  {journal} {\bibinfo  {journal} {International Journal of Modern Physics D}\
  }\textbf {\bibinfo {volume} {26}},\ \bibinfo {pages} {1750051} (\bibinfo
  {year} {2017})}\BibitemShut {NoStop}%
\bibitem [{\citenamefont {Liu}\ \emph {et~al.}(2017)\citenamefont {Liu},
  \citenamefont {Ding},\ and\ \citenamefont {Jing}}]{liu_effects_2017}%
  \BibitemOpen
  \bibfield  {author} {\bibinfo {author} {\bibfnamefont {C.-Q.}\ \bibnamefont
  {Liu}}, \bibinfo {author} {\bibfnamefont {C.-K.}\ \bibnamefont {Ding}}, \
  and\ \bibinfo {author} {\bibfnamefont {J.-L.}\ \bibnamefont {Jing}},\ }\href
  {\doibase 10.1088/0256-307X/34/9/090401} {\bibfield  {journal} {\bibinfo
  {journal} {Chinese Physics Letters}\ }\textbf {\bibinfo {volume} {34}},\
  \bibinfo {pages} {090401} (\bibinfo {year} {2017})}\BibitemShut {NoStop}%
\bibitem [{\citenamefont {Haroon}\ \emph {et~al.}(2019)\citenamefont {Haroon},
  \citenamefont {Jamil}, \citenamefont {Jusufi}, \citenamefont {Lin},\ and\
  \citenamefont {Mann}}]{haroon_shadow_2019}%
  \BibitemOpen
  \bibfield  {author} {\bibinfo {author} {\bibfnamefont {S.}~\bibnamefont
  {Haroon}}, \bibinfo {author} {\bibfnamefont {M.}~\bibnamefont {Jamil}},
  \bibinfo {author} {\bibfnamefont {K.}~\bibnamefont {Jusufi}}, \bibinfo
  {author} {\bibfnamefont {K.}~\bibnamefont {Lin}}, \ and\ \bibinfo {author}
  {\bibfnamefont {R.~B.}\ \bibnamefont {Mann}},\ }\href {\doibase
  10.1103/PhysRevD.99.044015} {\bibfield  {journal} {\bibinfo  {journal}
  {Physical Review D}\ }\textbf {\bibinfo {volume} {99}},\ \bibinfo {pages}
  {044015} (\bibinfo {year} {2019})}\BibitemShut {NoStop}%
\bibitem [{\citenamefont {Kimpson}\ \emph {et~al.}(2019)\citenamefont
  {Kimpson}, \citenamefont {Wu},\ and\ \citenamefont
  {Zane}}]{kimpson_spatial_2019}%
  \BibitemOpen
  \bibfield  {author} {\bibinfo {author} {\bibfnamefont {T.}~\bibnamefont
  {Kimpson}}, \bibinfo {author} {\bibfnamefont {K.}~\bibnamefont {Wu}}, \ and\
  \bibinfo {author} {\bibfnamefont {S.}~\bibnamefont {Zane}},\ }\href {\doibase
  10.1093/mnras/stz138} {\bibfield  {journal} {\bibinfo  {journal} {Monthly
  Notices of the Royal Astronomical Society}\ }\textbf {\bibinfo {volume}
  {484}},\ \bibinfo {pages} {2411} (\bibinfo {year} {2019})}\BibitemShut
  {NoStop}%
\bibitem [{\citenamefont {Babar}\ \emph {et~al.}(2020)\citenamefont {Babar},
  \citenamefont {Babar},\ and\ \citenamefont
  {Atamurotov}}]{babar_optical_2020}%
  \BibitemOpen
  \bibfield  {author} {\bibinfo {author} {\bibfnamefont {G.~Z.}\ \bibnamefont
  {Babar}}, \bibinfo {author} {\bibfnamefont {A.~Z.}\ \bibnamefont {Babar}}, \
  and\ \bibinfo {author} {\bibfnamefont {F.}~\bibnamefont {Atamurotov}},\
  }\href {\doibase 10.1140/epjc/s10052-020-8346-3} {\bibfield  {journal}
  {\bibinfo  {journal} {The European Physical Journal C}\ }\textbf {\bibinfo
  {volume} {80}},\ \bibinfo {pages} {761} (\bibinfo {year} {2020})}\BibitemShut
  {NoStop}%
\bibitem [{\citenamefont {Junior}\ \emph {et~al.}(2020)\citenamefont {Junior},
  \citenamefont {Crispino}, \citenamefont {Cunha},\ and\ \citenamefont
  {Herdeiro}}]{junior_spinning_2020}%
  \BibitemOpen
  \bibfield  {author} {\bibinfo {author} {\bibfnamefont {H.~C. D.~L.}\
  \bibnamefont {Junior}}, \bibinfo {author} {\bibfnamefont {L.~C.~B.}\
  \bibnamefont {Crispino}}, \bibinfo {author} {\bibfnamefont {P.~V.~P.}\
  \bibnamefont {Cunha}}, \ and\ \bibinfo {author} {\bibfnamefont {C.~A.~R.}\
  \bibnamefont {Herdeiro}},\ }\href {\doibase 10.1140/epjc/s10052-020-08572-w}
  {\bibfield  {journal} {\bibinfo  {journal} {The European Physical Journal C}\
  }\textbf {\bibinfo {volume} {80}},\ \bibinfo {pages} {1036} (\bibinfo {year}
  {2020})}\BibitemShut {NoStop}%
\bibitem [{\citenamefont {Bad\'\i{}a}\ and\ \citenamefont
  {Eiroa}(2021)}]{Badia:2021}%
  \BibitemOpen
  \bibfield  {author} {\bibinfo {author} {\bibfnamefont {J.}~\bibnamefont
  {Bad\'\i{}a}}\ and\ \bibinfo {author} {\bibfnamefont {E.~F.}\ \bibnamefont
  {Eiroa}},\ }\href@noop {} {\  (\bibinfo {year} {2021})},\ \Eprint
  {http://arxiv.org/abs/2106.07601} {arXiv:2106.07601 [gr-qc]} \BibitemShut
  {NoStop}%
\bibitem [{\citenamefont {Bisnovatyi-Kogan}\ and\ \citenamefont
  {Tsupko}(2017{\natexlab{b}})}]{Kogan:2017}%
  \BibitemOpen
  \bibfield  {author} {\bibinfo {author} {\bibfnamefont {G.~S.}\ \bibnamefont
  {Bisnovatyi-Kogan}}\ and\ \bibinfo {author} {\bibfnamefont {O.~Y.}\
  \bibnamefont {Tsupko}},\ }\href {\doibase 10.3390/universe3030057} {\bibfield
   {journal} {\bibinfo  {journal} {Universe}\ }\textbf {\bibinfo {volume} {3}}
  (\bibinfo {year} {2017}{\natexlab{b}}),\ 10.3390/universe3030057}\BibitemShut
  {NoStop}%
\bibitem [{\citenamefont {Atamurotov}\ \emph
  {et~al.}(2015{\natexlab{b}})\citenamefont {Atamurotov}, \citenamefont
  {Ahmedov},\ and\ \citenamefont {Abdujabbarov}}]{Atamurotov:2015}%
  \BibitemOpen
  \bibfield  {author} {\bibinfo {author} {\bibfnamefont {F.}~\bibnamefont
  {Atamurotov}}, \bibinfo {author} {\bibfnamefont {B.}~\bibnamefont {Ahmedov}},
  \ and\ \bibinfo {author} {\bibfnamefont {A.}~\bibnamefont {Abdujabbarov}},\
  }\href {\doibase 10.1103/PhysRevD.92.084005} {\bibfield  {journal} {\bibinfo
  {journal} {Physical Review D}\ }\textbf {\bibinfo {volume} {92}},\ \bibinfo
  {pages} {084005} (\bibinfo {year} {2015}{\natexlab{b}})}\BibitemShut
  {NoStop}%
\bibitem [{\citenamefont {Perlick}\ and\ \citenamefont
  {Tsupko}(2017{\natexlab{b}})}]{Perlick:2017}%
  \BibitemOpen
  \bibfield  {author} {\bibinfo {author} {\bibfnamefont {V.}~\bibnamefont
  {Perlick}}\ and\ \bibinfo {author} {\bibfnamefont {O.~Y.}\ \bibnamefont
  {Tsupko}},\ }\href {\doibase 10.1103/PhysRevD.95.104003} {\bibfield
  {journal} {\bibinfo  {journal} {Physical Review D}\ }\textbf {\bibinfo
  {volume} {95}},\ \bibinfo {pages} {104003} (\bibinfo {year}
  {2017}{\natexlab{b}})}\BibitemShut {NoStop}%
\bibitem [{\citenamefont {Carter}(1968)}]{Carter:1968}%
  \BibitemOpen
  \bibfield  {author} {\bibinfo {author} {\bibfnamefont {B.}~\bibnamefont
  {Carter}},\ }\href {\doibase 10.1103/PhysRev.174.1559} {\bibfield  {journal}
  {\bibinfo  {journal} {Physical Review}\ }\textbf {\bibinfo {volume} {174}},\
  \bibinfo {pages} {1559} (\bibinfo {year} {1968})}\BibitemShut {NoStop}%
\bibitem [{\citenamefont {Mino}(2003)}]{Mino:2003}%
  \BibitemOpen
  \bibfield  {author} {\bibinfo {author} {\bibfnamefont {Y.}~\bibnamefont
  {Mino}},\ }\href {\doibase 10.1103/PhysRevD.67.084027} {\bibfield  {journal}
  {\bibinfo  {journal} {Physical Review D}\ }\textbf {\bibinfo {volume} {67}},\
  \bibinfo {pages} {084027} (\bibinfo {year} {2003})}\BibitemShut {NoStop}%
\bibitem [{\citenamefont {Misner}\ \emph {et~al.}(2017)\citenamefont {Misner},
  \citenamefont {Thorne},\ and\ \citenamefont {Wheeler}}]{Misner:1973}%
  \BibitemOpen
  \bibfield  {author} {\bibinfo {author} {\bibfnamefont {C.~W.}\ \bibnamefont
  {Misner}}, \bibinfo {author} {\bibfnamefont {K.~S.}\ \bibnamefont {Thorne}},
  \ and\ \bibinfo {author} {\bibfnamefont {J.~A.}\ \bibnamefont {Wheeler}},\
  }\href
  {https://press.princeton.edu/books/hardcover/9780691177793/gravitation}
  {\emph {\bibinfo {title} {Gravitation}}}\ (\bibinfo  {publisher} {Princeton
  University Press},\ \bibinfo {year} {2017})\BibitemShut {NoStop}%
\bibitem [{\citenamefont {Lancaster}\ and\ \citenamefont
  {Blundell}(2014)}]{Lancaster:2014}%
  \BibitemOpen
  \bibfield  {author} {\bibinfo {author} {\bibfnamefont {T.}~\bibnamefont
  {Lancaster}}\ and\ \bibinfo {author} {\bibfnamefont {S.~J.}\ \bibnamefont
  {Blundell}},\ }\href {\doibase 10.1093/acprof:oso/9780199699322.001.0001}
  {\emph {\bibinfo {title} {Quantum {Field} {Theory} for the {Gifted}
  {Amateur}}}}\ (\bibinfo  {publisher} {Oxford University Press},\ \bibinfo
  {year} {2014})\BibitemShut {NoStop}%
\bibitem [{\citenamefont {Schee}\ and\ \citenamefont
  {Stuchlík}(2009)}]{Schee:2009}%
  \BibitemOpen
  \bibfield  {author} {\bibinfo {author} {\bibfnamefont {J.}~\bibnamefont
  {Schee}}\ and\ \bibinfo {author} {\bibfnamefont {Z.}~\bibnamefont
  {Stuchlík}},\ }\href {\doibase 10.1142/S0218271809014881} {\bibfield
  {journal} {\bibinfo  {journal} {International Journal of Modern Physics D}\
  }\textbf {\bibinfo {volume} {18}},\ \bibinfo {pages} {983} (\bibinfo {year}
  {2009})}\BibitemShut {NoStop}%
\bibitem [{\citenamefont {Boyer}\ and\ \citenamefont
  {Lindquist}(1967)}]{Boyer:1967}%
  \BibitemOpen
  \bibfield  {author} {\bibinfo {author} {\bibfnamefont {R.~H.}\ \bibnamefont
  {Boyer}}\ and\ \bibinfo {author} {\bibfnamefont {R.~W.}\ \bibnamefont
  {Lindquist}},\ }\href {\doibase 10.1063/1.1705193} {\bibfield  {journal}
  {\bibinfo  {journal} {Journal of Mathematical Physics}\ }\textbf {\bibinfo
  {volume} {8}},\ \bibinfo {pages} {265} (\bibinfo {year} {1967})},\ \Eprint
  {http://arxiv.org/abs/https://doi.org/10.1063/1.1705193}
  {https://doi.org/10.1063/1.1705193} \BibitemShut {NoStop}%
\bibitem [{\citenamefont {Byrd}\ and\ \citenamefont
  {Friedman}(1971)}]{handbookElliptic}%
  \BibitemOpen
  \bibfield  {author} {\bibinfo {author} {\bibfnamefont {P.}~\bibnamefont
  {Byrd}}\ and\ \bibinfo {author} {\bibfnamefont {M.}~\bibnamefont
  {Friedman}},\ }\href
  {https://link.springer.com/book/10.1007/978-3-642-65138-0#about} {\emph
  {\bibinfo {title} {Handbook of elliptic integrals for engineers and
  scientists}}},\ Grundlehren der mathematischen Wissenschaften\ (\bibinfo
  {publisher} {Springer-Verlag},\ \bibinfo {year} {1971})\BibitemShut {NoStop}%
\end{thebibliography}%

\end{document}